%% file: main_BioEpi.tex
\documentclass[]{interact}

\usepackage[authoryear,round,numbers,sort&compress]{natbib}
\bibpunct[, ]{[}{]}{,}{n}{,}{,}
\renewcommand\bibfont{\fontsize{10}{12}\selectfont}
\makeatletter
\def\NAT@def@citea{\def\@citea{\NAT@separator}}
\makeatother
\usepackage{times}
\usepackage{w-thm}
\usepackage{verbatim,color,amssymb}
\usepackage{graphicx}
\usepackage{epstopdf}
\usepackage{epsfig}
\usepackage{mathtools}
\usepackage{amssymb,amsthm}
\usepackage{textcomp}
\usepackage{amsmath}
\usepackage{rotating}
\usepackage{makecell}
\usepackage{array}
\usepackage{longtable}

\newcommand{\bx}{{\bf x}}

\newcommand{\bz}{{\bf z}}

\newcommand{\be}{{\bf e}}

\newcommand{\bgamma}{\mbox{\boldmath $\Gamma$}}

\newcommand{\bomega}{\mbox{\boldmath $\Omega$}}

\theoremstyle{plain}

\theoremstyle{definition}

\usepackage[]{graphicx}
\chardef\bslash=`\\ 

\def\S{{\cal D}}

\def\bse{\begin{eqnarray*}}
\def\ese{\end{eqnarray*}}
\def\be{\begin{eqnarray*}}
\def\ee{\end{eqnarray*}}
\def\bq{\begin{equation}}
\def\eq{\end{equation}}
\def\bse{\begin{eqnarray*}}
\def\ese{\end{eqnarray*}}

\def\trans{^{\rm T}}

\def\bU{{\mathbf U}}
\def\bZ{{\mathbf Z}}
\def\bW{{\mathbf W}}

\def\bX{{\mathbf X}}
\def\bz{{\mathbf z}}

\newcommand{\bTheta}{\mbox{\boldmath $\Theta$}}

\graphicspath{{./figs/}}

\begin{document}

\title{Adjusting for bias due to measurement error in functional quantile regression models with error-prone functional and scalar covariates}

\author{
\name{Xiwei Chen\textsuperscript{a}, Yuanyuan Luan\textsuperscript{a}, Roger S. Zoh\textsuperscript{a}, Lan Xue\textsuperscript{b}, Sneha Jadhav\textsuperscript{c}, and Carmen D. Tekwe\textsuperscript{a}\thanks{CONTACT Carmen D. Tekwe. Email: ctekwe@iu.edu}}
\affil{\textsuperscript{a}Department of Epidemiology and Biostatistics, Indiana University, Bloomington, Indiana, US; \textsuperscript{b}Department of Statistics, Oregon State University, Corvallis, Oregon, US; \textsuperscript{c}Department of Mathematics and Statistics, Wake Forest University, Winston-Salem, North Carolina, US}
}

\maketitle

\begin{abstract}
Wearable devices enable the continuous monitoring of physical activity (PA) but generate complex functional data with poorly characterized errors. Most work on functional data views the data as smooth, latent curves obtained at discrete time intervals with some random noise with mean zero and constant variance. Viewing this noise as homoscedastic and independent ignores potential serial correlations. Our preliminary studies indicate that failing to account for these serial correlations can bias estimations. In dietary assessments, epidemiologists often use self-reported measures based on food frequency questionnaires that are prone to recall bias. With the increased availability of complex, high-dimensional functional, and scalar biomedical data potentially prone to measurement errors, it is necessary to adjust for biases induced by these errors to permit accurate analyses in various regression settings. However, there has been limited work to address measurement errors in functional and scalar covariates in the context of quantile regression. Therefore, we developed new statistical methods based on simulation extrapolation (SIMEX) and mixed effects regression with repeated measures to correct for measurement error biases in this context. We conducted simulation studies to establish the finite sample properties of our new methods. The methods are illustrated through application to a real data set. 
\end{abstract}


\section{BACKGROUND AND MOTIVATING EXAMPLE}\label{sec:intro}

\subsection{Background on Models}
Quantile regression is a robust statistical technique that permits assessment of intervention and covariate effects on the median or other quantile functions of a response \citep{koenker2005quantile}. While mean regression models focus on modeling covariate effects on mean responses, quantile regression can be applied to examine covariate effects on the entire distribution of a response. Thus, quantile regression methods can be used to investigate covariate effects beyond the mean. These methods have been applied in studies of obesity where research interests may lie in determining intervention effects on individuals whose body weights may fall in the upper tail of body weight distributions \citep{bottai2014use,tekwe2022estimation}. Such individuals are also at higher risks for obesity and its related health outcomes when compared to individuals with average body weights. Quantile regression has been applied widely to other areas, including investigation of factors involved with cardiovascular disease \citep{degerud2015vitamin,blankenberg2016troponin,ng2014physical,beets2016we,beets2016physical}. However, to date, most applications of quantile regression have been used with cross-sectional data with scalar-valued covariates.

Most approaches to estimate regression models with functional covariates address the estimation problem from a functional data perspective \citep{chen2012,goldsmith2011penalized, goldsmith2011functional,greven2011longitudinal,yao2005functional,cardot2007smoothing,james2002generalized,crambes2009smoothing,tekwe2017functional}. Because functional covariates are considered latent and measured by smooth but noisy curves \citep{silverman2005}, the resulting white noise in the functional covariates is assumed to have an independent error structure, even though the covariates are often observed serially with complex heteroscedastic errors which may affect the validity of the data analysis. Greven et al. \citep{greven2011longitudinal} developed the longitudinal functional principal components analysis (LFPCA) to analyze functional data collected at multiple time points. In the LFPCA methods, the functional data were decomposed into a time-dependent mean component, and subject-specific variability, and extended FPCA methods were used for dimension reduction. The functional data error terms were assumed to be white noise with a constant variance, $\sigma^{2}$. In Tekwe et al. \citep{tekwe2017functional}, we also developed FPCA-based methods to correct for measurement errors associated with multiple functional data obtained from devices for energy expenditure. Using a two-stage estimation approach based on the expectation-maximization(EM) algorithm, we assumed a homoscedastic covariance matrix for the measurement errors. Functional additive mixed models have also been proposed, however, these models focus on functional responses that may be correlated, which is a different setting from regression settings with measurement error-prone functional covariates \citep{scheipl2015functional,scheipl2016generalized}.

Some recent advances have been made to correct for biases due to measurement error with arbitrary heteroscedastic covariance structures associated with functional covariates in classical regression models \citep{crainiceanu2009generalized,chakraborty2017regression,tekwesim,jadhav2022function,zoh2022fully}. For example, Crainiceanu et al. \citep{crainiceanu2009generalized} developed the generalized multilevel functional regression to model exposures with a multilevel functional structure. They developed the multilevel FPCA to address the functional analog of exposure measurement error models. Chakraboty and Panaretos \citep{chakraborty2017regression} proposed calibration-based nonparametric functional regression with functional measurement error. In Tekwe et al. \citep{tekwesim}, a generalized method of moments approach was developed to estimate the scalar on a function regression model with measurement error. Using an instrumental variable, we estimated the covariance matrix associated with the measurement error under the assumption that the error structures were non-white noise. Jadhav and colleagues \citep{jadhav2022function} proposed a regression calibration approach using instrumental variables that allows unrestricted functional measurement error. Finally, Zoh et al. \citep{zoh2022fully} recently introduced a Bayesian approach to the functional linear regression model with complex heteroscedastic measurement errors. 


While work on measurement error correction with functional covariates in regression models is growing, limited work has been done to address measurement errors associated with both functional and scalar covariates in regression models. In this manuscript, we consider developing approaches for measurement error correction associated with both functional and scalar covariates prone to measurement errors in quantile regression models. \textcolor{black}{Using repeated measures as identifying information to estimate the covariance functions associated with the measurement errors, we propose simulation extrapolation (SIMEX) and two fast and scalable approaches based on extensions of mixed effects model for bias correction due to measurement error in multi-level high dimensional settings.}

\subsection{Motivating Example}
Obesity development has been linked to a chronic imbalance between energy expenditure and energy intake. Energy expenditure is difficult to measure directly in large epidemiologic studies; therefore, researchers often estimate it from physical activity \citep{hallal2013energy}. Research-grade body-worn monitoring devices are the most objective way to measure physical activity in community-dwelling individuals. However, the accuracy of data generated by such devices may be questionable \citep{shcherbina2017accuracy,reddy2018accuracy,jakicic1999accuracy,staudenmayer2012statistical,tekwesim}. For example, we recently demonstrated that device-measured physical activity is prone to high dimensional heteroscedastic measurement error, with poorly understood consequences for obesity studies \citep{tekwesim,tekwe2017functional}. In nutritional epidemiology, dietary assessments also suffer from measurement error. Because research participants typically recall or record their usual dietary intake retrospectively, the resulting data may reflect under- or over-reporting. For example, under-reporting of dietary intake in 24-hour recall and food frequency questionnaires can range from 10\% to 50\% \citep{subar2003using,park2007underreporting,macdiarmid1998assessing,mendez2011alternative}. We seek to determine how fiber intake and physical activity influence quantile functions of body mass index (BMI) in community-dwelling adults living in the United States. In our motivating example, we analyzed data from the National Health and Nutrition Examination Survey (NHANES) on device-measured physical activity and self-reported fiber intake as an indicator of dietary intake. We developed statistical models for correcting measurement errors in both scalar and functional covariates prone to measurement error. These models allow arbitrary heteroscedastic covariate error structures for the measurement error in the functional covariate and scalar error structures in the scalar covariate. We applied our developed error correction methods to the motivating example and evaluated their finite sample properties through extensive simulations. 

The manuscript is formatted as follows, we introduce our models in Section \ref{sec:models} and discuss estimation approaches in Section \ref{sec:estimate}. In Section \ref{subsec:iv}, we describe our SIMEX and functional mixed effects-based approaches. We present our simulation results and the application of our methods to our motivating example in Sections \ref{sec:simulation} and \ref{sec:application}, respectively. We end with some concluding remarks in Section \ref{sec:conclusion}.

\section{MODELS}\label{sec:models}
Let $\{Y_{i},X_{1i}(t),X_{2i},\bZ_i\}, i=1,\hdots,n$ represent a continuous scalar outcome $Y_i$, a random \textcolor{black}{square integrable} functional covariate $X_{1i}(t)$ defined on the unit interval, a continuous scalar covariate $X_{2i}$, and a $p \times 1$ vector of error-free covariates $\bZ_i$, respectively, for the $i${th} subject. We define $Q_{\tau}\{Y_{i}|X_{1i}(t),X_{2i},\bZ_i\}$ as $F^{-1}\{Y_{i}|X_{1i}(t),X_{2i},\bZ_i\}(\tau)$ and $F\{Y_{i}|X_{1i}(t),X_{2i},\bZ_i\} = P\{Y_{i} \leq y |X_{1i}(t),X_{2i},\bZ_i\}$, and propose the following conditional functional quantile regression model for $Y_i$:
 \begin{eqnarray}
 Q_{\tau}\{Y_{i}|X_{1i}(t),X_{2i},\bZ_i\} &=& \beta_{0\tau}+\int_{0}^{1} \beta_{1\tau}(t)X_{1i}(t)dt+ \beta_{2\tau}X_{2i} +\bZ_{i}\trans\bgamma_{\tau},  \label{eq1}  \\
   W_{1ij}(t) &=&  X_{1i}(t) + U_{1ij}(t), j=1,\hdots,J, \text{and} \label{eq2} \\
   W_{2il} &=&  X_{2i} + U_{2il}, l=1,\hdots,L. \label{eq3}
 \end{eqnarray}
Equations \ref{eq1} - \ref{eq3} describe the $i${th} participant at wear time $t$, the time at which the wearable device is worn. We re-parameterized wear time such that $t \in [0,1]$. Equation \ref{eq1} models the $\tau${th} conditional quantile function for the continuous scalar outcome, $Y_i$, or $BMI_i$ in our data application. The true unobserved functional covariate is $X_{1i} = \{X_{1i}(t) \in [0,1]\}$, while $X_{2i}$ is a scalar covariate prone to classical additive measurement error. The vector of error-free covariates, $\bZ_{i}$, includes $p$ error-free covariates such as age, sex, and race/ethnicity for the $i${th} participant. In our data application, we have proxy measures for both latent measures of physical activity intensity and fiber intake, respectively. In Equation \ref{eq2}, $W_{1ij}(t)$ is the device-measure of physical activity intensity for subject $i$ at wear time $t$ on day $j$, approximating the \textcolor{black}{usual patterns of true weekly physical activity intensity}, $X_{1i}(t)$, with some measurement error $U_{1ij}(t)$. The $W_{2il}$ is a scalar covariate for subject $i$ prone to error on day $l$. Also, $W_{2il}$ corresponds to self-reported responses in food frequency questionnaires which we use as a proxy for \textcolor{black}{usual dietary intake}, $X_{2i}$, with some measurement error $U_{2il}$. \textcolor{black}{We note here that the assumptions of fixed values for both $X_{1i}(t)$ and $X_{2i}$ across days indicate that interest in our estimation lies in how usual patterns of true weekly physical activity intensity and true usual measures fiber intake influence the outcomes. However, repeated measurements are obtained across multiple days to approximate these true latent measures to account for potential intra- and inter-day variability across the repeated measures.} In Equation \ref{eq1}, $\beta_{0\tau}$ is an intercept, $\beta_{1\tau}(t)$ is a functional-coefficient, $\beta_{2\tau}$ is a scalar coefficient representing the effects of the two covariates that are contaminated with additive measurement error, and $\bgamma_{\tau}$ represents a $p \times 1$ vector of coefficients associated with the error-free covariates for the $\tau${th} quantile. \textcolor{black}{We assume $\beta_{1\tau}(t)$ is a smooth function that can be well approximated by spline functions}, and that  $E\left\{U_{1ij}(t)|X_{1i}(t)\right\}=0$, $E(U_{2il}|X_{2i}) = 0$, and $cov\{U_{1ij}(t),U_{2il}\} = 0$ for all $i,j,l$ and $t$. Consequently, the surrogates $W_{1ij}(t)$ and $W_{2il}$ are unbiased measures for the true exposures $X_{1i}(t)$ and $X_{2i}$, respectively. \textcolor{black}{We also assume that the measurement errors are independent of the response and $X_{1i}(t)$ and $X_{2i}$}. In functional data analysis, it is often assumed that $U_{1ij}(t)$ is i.i.d with a common variance across time $t$. We instead consider an unstructured functional measurement error for $U_{1ij}(t)$.

The partially functional linear model in Equation \ref{eq1} is a useful generalization of both classical and functional linear quantile regression models, but existing methods often require specification of the conditional distributions of the response given the true functional covariates, which are usually unavailable in quantile regression models. Also, Equation \ref{eq1} involves a functional covariate of infinite dimension, which poses both computational and theoretical difficulties. We consider both SIMEX and more scalable approaches based on a functional mixed effects model to correct measurement errors.

\section{Estimation} \label{sec:estimate}
Prior to estimating the model parameters in Equations \ref{eq1} - \ref{eq3}, we first reduce the dimensions of the functional terms in the models through basis expansions. We assume the functional measurement error, $U_{1ij}(t)$, in Equation \ref{eq2} is a Gaussian process with zero mean and an unknown covariance function. For this estimation, we approximate $\beta_{1\tau}(t)$ in Equation \ref{eq1} with polynomial splines, writing $
\beta_{1\tau}(t)\approx \sum_{k=1}^{K_{n}}\omega_{ \tau k}b_{k}(t)$, where $\left\{\omega
_{\tau k}\right\}_{k=1}^{K_{n}}$ are unknown spline coefficients and $\left\{
b_{k}(t)\right\} _{k=1}^{K_{n}}$ are a set of spline basis functions on $%
[0,1]$. Let $X_{1ik}=\int_{0}^{1}X_{1i}(t)b_{k}(t)dt$, $W_{1ijk}=%
\int_{0}^{1}W_{1ij}(t)b_{k}(t)dt$, and $U_{1ijk}=\int_{0}^{1}U_{1ij}(t)b_{k}(t)dt$. Following spline approximations, the measurement error model in Equations \ref{eq1} - \ref{eq3} becomes
\begin{eqnarray}
Q_{\tau}\{Y_{i}|X_{1i}(t),X_{2i},\bZ_i\} &\approx& \beta_{0\tau}+\sum_{k=1}^{K_{n}}\omega_{k\tau}X_{1ik} +\beta_{2\tau}X_{2i}+ \bZ_{i}\trans\bgamma_{\tau}; \label{eq4}\\
W_{1ijk} &=&X_{1ik}+U_{1ijk}\hspace{0.5cm}k=1,...,K_{n}; \text{ and }  \label{eq5} \\
W_{2il} &=&  X_{2i} + U_{2il},  \label{eq6}
\end{eqnarray}
where $\left\{ U_{1ij1},\ldots ,U_{1ijK_{n}}\right\} $ are correlated errors. This reparameterized model is a multivariate linear quantile regression with measurement error. However, unlike the multivariate linear quantile regression with measurement error, the linear covariates in Equation \ref{eq4} and Equation \ref{eq5} are not fixed and increase with increasing sample size. In the absence of measurement error, the unknown coefficients in Equation \ref{eq4} can be estimated by minimizing
\begin{equation}
\tilde{{\bTheta}}_{\tau}=\text{argmin}_{{\bTheta}_{\tau}} \sum_{i=1}^{n}\rho (Y_{i},{\bX}_{1i},X_{2i},{\bZ}_i, {\bTheta}_{\tau}),\label{eq:true}
\end{equation}
where ${\bX}_{1i}=\left( X_{1i1},\cdots, X_{1iK_{n}}\right)^T$,
${\bTheta}_{\tau}= (\beta_{0\tau},\omega_{1\tau},\cdots,\omega_{K_n\tau}, \beta_{2\tau},
\bgamma_{\tau})^T$, and $\rho(y,{\bx_1}, \bx_2, {\bz},{\bTheta},\bTheta_{\tau})=\rho_{\tau}(y- \beta_{0\tau}-{\bf x_1}^{T}{\bf\omega_{\tau} }-{x_2}^T{\beta_{2\tau} }-{\bf z}^{T}{\bf\gamma_{\tau} })$, with $\rho_{\tau}(\epsilon)=\epsilon\left\{\tau-I(\epsilon<0)\right\}$ and $\bomega_{\tau} = (\omega_{1\tau}, \hdots, \omega_{K_n\tau})^T$.

\subsection{Assumptions}
Let $\bX_{1i}= (X_{1i1}, \hdots, X_{1iK_n})^T$, $\bW_{1ij}= (W_{1ij1}, \hdots, W_{1ijK_n})^T$, $\bU_{1ij}= (U_{1ij1}, \hdots, U_{1ijK_n})^T$, $\bW_{2i}= (W_{2i1}, \hdots, W_{2iL})^T$, and $\bU_{2i}= (U_{2i1}, \hdots, U_{2iL})^T$. We restate models and assumptions from Section \ref{sec:models} in terms of reparameterized models in Equations \ref{eq4} - \ref{eq6} as follows:

\begin{itemize}
\item[(i)] $\mathrm{Cov}(\bX_{1il},\bU_{1ijm}) =0$ and $\mathrm{Cov}(X_{2i},\bU_{2il})=0$. Thus, the covariances between paired observations are zero.



\item[(ii)] $\mathrm{Cov}(\bW_{1ij},\bU_{1ij}) \neq \b0$ and $\mathrm{Cov}(\bW_{1ij},\bX_{1i}) \neq \b0$. Similarly, $\mathrm{Cov}(\bW_{2i},\bU_{2i}) \neq \b0$ and $\mathrm{Cov}(\bW_{2i},X_{2i}) \neq \b0$.

\item[(iii)] $\mathrm{Cov}(U_{ik},U_{ij}) \neq 0$ for $k \ne j$.

\item[(iv)] $\bU_{1ij} \sim MVN(0,\Sigma_{u1})$ and $U_{2i} \sim N(0,\sigma_{u2}^{2})$.
\end{itemize}
Assumptions A1 represents the usual classical measurement error assumptions. Assumption A4 introduces normality on the measurement errors, which is necessary for our functional mixed effects models-based approaches. \textcolor{black}{However, our proposed SIMEX method is a semiparametric-based approach and remains valid for measurement error correction when the normality assumption is violated.}

\section{Approaches to correcting measurement error}\label{subsec:iv}
Replacing the true unobservable covariates with their observed measurement error-prone surrogates leads to estimation bias \citep{Fuller1987,Carroll:2006}, but the direction of the bias depends on the model. Correcting biases due to measurement errors in covariates in regression models often requires additional information on the surrogate or observed measures. The additional information includes replicates or other data on the surrogate measures that allow estimating the measurement error variance and covariance functions. Instrumental variables, external validation data, or assumptions of known measurement error variances and covariance functions may also be used as additional identifying information. We estimate this covariance function with repeated measurements or replicates on the functional surrogate measures for the functional measurement error in Equation \ref{eq2} and replicates on the scalar-valued surrogate measure are also used to obtain estimates of the variance term for the measurement error in Equation \ref{eq3}. We developed error correction methods based on mixed effects and SIMEX approaches that reduce biases when estimating the associations between the error-prone covariates with quantile levels of the response.

\subsection{Scalable mixed effects-based approaches}
Researchers frequently use regression calibration \citep{carroll1990approximate} to correct measurement errors in settings such as radiation epidemiology \citep{little2008new,tekwe:2014,heidenreich2010use,nakashima2019impact} and nutritional epidemiology \citep{spiegelman1997regression,bennett2017systematic}. Regression calibration requires either a gold standard for the true unobserved covariate, replicates, or an instrumental variable \textcolor{black}{\citep{Fuller1987,Carroll:2006,spiegelman1997regression}}. It is an approximation-based approach for obtaining re-calibrated measures of the true unobserved exposures by using additional identifying data. To do this in the classical additive measurement error models in Equations \ref{eq2} and \ref{eq3}, the conditional distribution of the true latent covariate given the observed data are approximated with additional identifying data. Once approximated, the latent true covariates are replaced with the means of the conditional distribution. \textcolor{black}{This approach works well with scalar-valued covariates prone to errors, however, its extension to the case of multi-level functional covariates is not trivial. Functional covariates are not mere generalizations of scalar covariates since the functions are infinite dimensional objects making extensions of regression calibration approaches to this setting more complex. Some of the challenges with this approach include the need to evaluate integrals that may be intractable and the need to invert high dimensional covariance functions. While approximation methods exist to approximate multi-dimensional intractable integrals, these methods can be slow to converge and computationally intensive.}

For longitudinally observed error-prone surrogate measures, linear mixed-effects models-based approaches have been implemented for measurement error correction \citep{strand2014regression}. This approach to measurement error correction may be performed using standard statistical packages for mixed effects models and does not require simulating directly from the conditional distribution of the true covariate given the observed data. Instead, replicates of the longitudinal error-prone surrogate are used in a calibration model for the true covariate. In practice, under the assumption that the repeatedly observed surrogate measure is unbiased for the true unobserved covariate, the predicted values from the regression of the surrogate on the intercept in a linear mixed effects model may be used as measurement error-corrected measures for the true covariate when assessing its association with the outcome of interest. For example, Strand and colleagues \citep{strand2014regression} demonstrated the use of mixed effects models for measurement error correction in longitudinal data with repeatedly observed exposures prone to measurement error.

For the mixed effects approaches to measurement error correction, we view the classical measurement error models as linear mixed effects models and propose a two-stage approach for the measurement error correction. In the first stage, we use mixed effects-based methods to obtain error-corrected measures for the true covariates. The second stage involves plugging the error-corrected surrogates for the true covariates in the regression model for the response. We describe the steps below:

\begin{enumerate}
\item[(i)] For any $t\in [0,1]$, the mixed effects model for repeated functional surrogate measures is
\begin{eqnarray*}
 W_{1ij}(t) &=& \alpha_{0}(t) + \alpha_{0i}(t) + \varepsilon_{1ij}(t)
 \text{ and}  \label{RC_Wt}
\end{eqnarray*}
While the model for the scalar covariate prone to error is
\begin{eqnarray*}
W_{2il} &=& \eta_{0} + \eta_{0i} + \varepsilon_{2il},
\end{eqnarray*}
where $\alpha_{0}(t)$ and $\eta_{0}$ are the fixed intercepts. The random intercepts, $\alpha_{0i}(t)$ and $\eta_{0i}$, and the model errors, $\varepsilon_{1ij}(t)$ and $\varepsilon_{2il}$, are centered Gaussian errors, and the functional terms $\alpha_{0i}(t)$ and $\varepsilon_{1ij}(t)$ have arbitrary unknown correlation structures. In addition, we assume they are independent and identically distributed across subjects. $\alpha_{0i}(t)$ and $\eta_{0i}$ are subject-specific random intercepts. The mixed effects models approach does not require reducing the dimension of the functional terms in Equation \ref{RC_Wt} to obtain the predicted values of the true covariates that will be used as re-calibrated measures for $X_{1i}(t)$ in the regression models for the response. 
   
We propose two distinct mixed effects-based methods for the measurement error correction: fast univariate inference (FUI) \citep{cui2022fast} and fast short multivariate inference (FSMI) \textcolor{blue}{xxx et al. (manuscript under review)}. In both methods, we obtain the predicted values for the true measures as $\hat{X}_{1i}(t) = \hat{\alpha}_{0}(t) + \hat{\alpha}_{0i}(t)$ and $\hat{X}_{2i} = \hat{\eta}_{0} + \hat{\eta}_{0i}$. We detail these methods further in the next paragraph.

\item[(ii)] For the functional error-prone covariate, we reduce the dimension of $\widehat{X}_{1i}(t)$ with spline basis expansion in the second stage of this approach as described in Section\ref{sec:estimate}, and obtain $\widehat{\bX}_{1i}=\left(\widehat{X}_{1i1},\cdots, \widehat{X}_{1iK_n}\right)^{T}$.

\item[(iii)] Next, we minimize the check function in Equation \ref{eq:true} by replacing $\bX_{1i}$ and $X_{2i}$ with $\widehat{\bX}_{1i}$ and $\widehat{X}_{2i}$, respectively, to obtain
$\widehat{\bTheta}_{\tau}=\{\widehat{\beta}_{0\tau},\widehat{\omega}_{1\tau},\cdots,\widehat{\omega}_{K_n\tau},\widehat{\beta}_{2\tau},\widehat{{\bf\gamma}}_{\tau}\}$. Then estimate the functional coefficient in Equation \ref{eq1} as $\widehat{\beta}_{1\tau}(t)=\sum_{k=1}^{K_n} \widehat{\omega}_{k\tau}b_k(t)$.
\end{enumerate}

This approach allows the measurement error function to have an arbitrary covariance structure, such as unstructured (UN), compound symmetry (CS), or autoregressive with lag 1, i.e., $AR(1)$. The FUI method was proposed by Cui et al (2022) as a fast and scalable approach to estimating multi-level generalized functional linear regression models \citep{cui2022fast}. While fitting multi-level functional data can be computationally intensive, the FUI method simplifies the high-dimensional estimation by performing the analysis by time. The term "univariate" in FUI refers to the point-wise fitting approach of the multi-level functional data for each wear time separately. In particular, FUI fits a point-wise linear mixed effects model for each wear time $t, t= 1, \hdots, T$ across all $J$ days. For each $t$, the predicted value of the true measure is $\hat{X}_{1i}(t) = \hat{\alpha}_{0}(t) + \hat{\alpha}_{0i}(t)$, resulting in $T$ predicted values treated as functional data.

Because FUI fits the functional data by each $t$ across multiple days, it might not account for the serial correlation across different wear times in our motivating example. To address this potential drawback, we developed the FSMI approach \textcolor{blue}{xxx et al. (manuscript under review)}, which also fits point-wise linear mixed effects models to the data. However, FSMI does this by using multiple wear times concurrently in computing a moving average across all the $T$ wear times. The term "multivariate" in FSMI refers to using multiple wear times concurrently when fitting the point-wise linear mixed effects model. Both FUI and FSMI methods are parametric approaches that permit the specification of the error structure of the measurement error covariance matrix. We obtained the standard errors of the estimated parameters with $95\%$ nonparametric bootstrap confidence intervals.

Because the scalar covariate in our motivating example was measured at a single time point across multiple days, we use a linear mixed effects model to obtain the predicted values for $X_{2i}$. The standard errors for the scalar covariates were also obtained with $95\%$ nonparametric bootstrap confidence intervals.


\subsection{Simulation Extrapolation}
The SIMEX \citep{cookStefanski1994} algorithm was designed for additive measurement error models when additional identifying information is available in the observed data for estimating the variance of the measurement error. SIMEX is also applicable when the measurement error generating process can be simulated with Monte Carlo methods \citep{Carroll:2006}. SIMEX works by adding measurement errors to the observed values of the true covariates, $\bW_{1i}$ and $W_{2i}$, through re-sampling in the simulation step. After this re-sampling step, the model parameters are estimated and a trend of the measurement error as a function of the estimated parameters is established. The simulation and estimation steps are repeated $S$ times\citep{cookStefanski1994}. 
The final stage of SIMEX involves extrapolating the established trends back to the case of no measurement error. In our use of SIMEX, we assume that the measurement errors in Equations \ref{eq2} and \ref{eq3} are additive where paired terms $\bX_{1i}$ and $\bU_{1i}$ and $X_{2i}$ and $U_{2i}$ are mutually independent. We also assume $\bU_{1i}$ is independent of $(Y_i, \bX_{1i}, X_{2i}, \bZ_{i},U_{2i})$ and $U_{2i}$ is independent of $(Y_i, \bX_{1i}, X_{2i}, \bZ_{i},\bU_{1i})$. Previously we described SIMEX for measurement error in a quantile regression model with a functional covariate prone to measurement error where a functional instrumental variable is used for model identification \citep{tekwe2022estimation}. In our current implementation, we use SIMEX for scalar and functional covariates prone to measurement errors where repeated measures are available for estimating the measurement error covariances. There are ten steps in this SIMEX process:

\begin{enumerate}
\item[(i)] Let $\overline{W_{1i \bullet}} (t)=\sum_{j=1}^{J}W_{1ij}(t)/J$ be the average functional covariate across replicates, and $\overline{W_{1\bullet \bullet}} (t)=\sum_{i=1}^{n}\sum_{j=1}^{J}W_{1ij}(t)/nJ$ be the overall average functional covariate. Obtain the basis expansion for $W_{1ij}(t)$, $\overline{W_{1i \bullet}} (t)$, and $\overline{W_{1\bullet \bullet}} (t)$ to arrive at ${W_{1ijk}}$, $\overline{W_{1i \bullet k}}$, and $\overline{W_{1\bullet \bullet k}}$, respectively, for $k= 1, \hdots, K_n$.

\item[(ii)] Let $A(i,j)$ represent an element in the $i${th} row and $j${th} column of a matrix $A$. Using the repeated measures  ${W_{1ijk}}$ and $W_{2il}$, estimate $\Sigma_{u1}$ and $\sigma_{u2}^{2}$ as
    \[
    \widehat{\Sigma}_{u1}(k,k^{\prime}) = \frac{\sum_{i=1}^{n} \sum_{j=1}^{J} \sum_{k=1}^{K_{n}} \left(W_{1ijk} - \overline{W_{1i\bullet  k}}\right)\left(W_{1ijk^{\prime}} - \overline{W_{1i\bullet k^{\prime}}}\right)}{n(J-1)K_n} \hspace{0.2cm}  \text{and}
    \]
    \[
    \widehat{\sigma}_{u2}^{2} = \frac{\sum_{i=1}^{n} \sum_{l=1}^{L}\left( W_{2il} - \overline{{W_{2i\bullet}}}\right)^{2}}{n(L-1)}.
    \]
$\widehat{\Sigma}_{u1}$ is a $K_n \times K_n$ covariance matrix, while $\widehat{\sigma}_{u2}^{2}$ is scalar.

\item[(iii)] Identify a sequence of increasing small positive numbers such as $\lambda = 0, 1/4, 1/2, \hdots,2$.

\item[(iv)] Simulate $\overline{\bU}_{1i} \sim MVN(0,\widehat{\Sigma}_{u1}/J)$ and $\overline{U}_{2i} \sim N(0,\hat{\sigma}_{u2}^{2}/L)$.

\item[(v)] Obtain the average surrogate measures across replicates $\overline{W_{1i\bullet k}}$ and $\overline{W_{2i}}$, respectively.

\item[(vi)] Obtain the re-measurements of the error-prone covariates as
\begin{eqnarray}
   \overline{W_{1i\bullet k}}(\lambda) &=& \overline{W_{1i\bullet k}} + \sqrt{\lambda}\overline{U}_{1ik} \text{ and }\\
   \overline{W_{2i}}(\lambda) &=& \overline{W_{2i\bullet}} + \sqrt{\lambda}\overline{U}_{2i}.
\end{eqnarray}

\item[(vii)] At each value of $\lambda$, minimize the check function in Equation \ref{eq:true} or perform the quantile regression at the quantile level of interest by replacing $\bX_{1i}=\left( X_{1i1},\cdots, X_{1iK_{n}}\right)^T$
with $\overline{\bW_{1i \bullet}}= (\overline{W_{1i \bullet 1}},\hdots, \overline{W_{1i \bullet K_n}})$ and $X_{2i}$ with $\overline{W_{2i}}$ for the $\tau$th quantile level of the response.

\item[(viii)] Repeat steps $4-7$ $S$ times. For the collection of all estimated coefficients in $\widehat{\bTheta}_{s}(\tau,\lambda)$ from Equation \ref{eq4} in the $s$th iteration for $s = 1,\hdots,S$, calculate
\begin{eqnarray}
   \widehat{\bTheta}_{\tau}(\lambda)&=& \frac{1}{S}\sum_{s=1}^{S}\widehat{\bTheta}_{s}(\tau,\lambda)
\end{eqnarray}
    over all $\lambda$ and all $S$ iterations to complete the simulation stage.

\item[(ix)] The extrapolation stage in our implementation involves establishing a quadratic extrapolation function for the relationship between the estimated coefficients, $\widehat{\bTheta}_{\tau}(\lambda)$ and $\lambda$.  In particular, let $\widehat{\theta}_{\delta \tau }(\lambda_m)$ be the $\delta$th element of $\widehat{\Theta}_{\tau}(\lambda_m)$. Then use
$\widehat{\theta}_{\delta \tau }(\lambda_m)  = \phi_{0\tau} + \phi_{1\tau}\lambda_m + \phi_{2\tau}\lambda_m^{2} + \varepsilon_m$ for $\{\lambda_m,\widehat{\theta}_{\delta \tau }(\lambda_m)\}_{m=0}^{M}$. The extrapolation function evaluated at $\lambda = -1$ and $\lambda=0$ yields the SIMEX and naive estimators, respectively.

\item[(x)] The SIMEX estimators for the functional, scalar, and error-free covariates are
 \begin{eqnarray}
 \widehat{\beta}_{1\tau,SIMEX}(t) &=& \sum_{k=1}^{K_{n}}\widehat{\omega}_{k\tau}(\lambda = -1)b_{k}(t),\\
  \widehat{\beta}_{2\tau,SIMEX}  &=& \widehat{\beta}_{2\tau}(\lambda = -1), \text{ and}\\
  \widehat{\gamma}_{p\tau,SIMEX}  &=& \widehat{\bgamma}_{p\tau}(\lambda = -1) \hspace{0.5cm} p=1,\hdots,P,
  \end{eqnarray}
  respectively. The corresponding naive estimators are
   \begin{eqnarray}
 \widehat{\beta}_{1\tau,Naive}(t)  &=& \sum_{k=1}^{K_{n}}\widehat{\omega}_{k\tau}(\lambda = 0)b_{k}(t),\\
  \widehat{\beta}_{2\tau,Naive}  &=& \widehat{\beta}_{2\tau}(\lambda = 0), \text{ and}\\
  \widehat{\gamma}_{p\tau,Naive} &=& \widehat{\bgamma}_{p\tau}(\lambda = 0) \hspace{0.5cm} p=1,\hdots,P.
  \end{eqnarray}

\end{enumerate}
We also computed $95\%$ nonparametric bootstrap confidence intervals for the SIMEX estimators.

\section{SIMULATION EXPERIMENTS} \label{sec:simulation}
We conducted simulations to investigate the finite sample properties of our methods in six different sets of simulations with varying conditions. In each set of simulations, for a uniform grid of points, \textcolor{black}{$t_1, \hdots, t_{100} \in [0,1]$}, we simulated the true functional covariate with $X_{1}(t_\delta)=1/[1+exp\{4*2*(t_\delta-0.5)\}] + \varepsilon_{X_1}(t_\delta)$, \textcolor{black}{$\delta = 1, \hdots, 100 \in [0,1]$}. We simulated the observed measures for $X_{1}(t_\delta)$ with $W_{1j}(t_\delta)=X_1(t_\delta)+U_{1j}(t_\delta), j=1,\hdots, 7$, where $\{\varepsilon_{X_1}(t_1), \hdots, \varepsilon_{X_1}(t_{100})\}^T  \sim MVN(0,\sigma_{X_1}^2)$. We simulated the true scalar covariate with $X_{2i} = 2 + \varepsilon_{X_2}$ and the scalar covariate prone to measurement error with $W_{2il}=X_2+U_2, l=1,..., 7$, where $\varepsilon_{X_2} \sim N(0, \sigma_{X_2}^{2})$. We simulated two scalar error-free covariates as $Z_{c} \sim N(1, \sigma_{Z_c}^{2})$ and $Z_{b} \sim Binomial(n, p_{Z_b})$. Finally, we generated the centered scalar response, $Y$, from a functional linear regression model with
\[Y = \int_{0}^{1} \beta_1(t)X_{1}(t)\,dt+\beta_2X_{2}+\gamma_1Z_{c}+\gamma_2Z_{b}+\varepsilon \sim N(0,0.1^2),\] where the functional coefficient for the true covariate $X_{1}(t)$ was $\beta_1(t) = sin(2\pi t)$. We generated all data in our simulations independently and performed the simulations at quantile levels of $\tau \in (0.25,0.50, 0.75,0.95)$. \textcolor{black}{The number of knots was selected by the Bayesian information criterion (BIC) method.} 


In the first simulation study, we examined the effects of varying sample sizes on $\widehat{\beta}_{1\tau}(t)$. We restricted the structure of the covariance matrices for the functional terms to be \{AR(1)\} with $\rho_{X_1} =\rho_{U_1} = 0.5$, $\sigma_{X_1}(t) = 3$, $U_{1}(t) \sim MVN\{0, \sigma_{U_1}(t)^2\}$ with $\sigma_{U_1}(t)= 2.5$, $\sigma_{X_2}=0.5$, $U_{2} \sim N(0, \sigma_{U_2}^{2})$ with $\sigma_{U_2}=0.25$, $\beta_2=0.5$, $\sigma_{Z_c}=0.5$, and $p_{Z_b}=0.6$. We considered sample sizes of $n=(100, 500, 1000, 2000, 5000)$.

\textcolor{black}{In the second simulation study, we investigated the effects of distributional assumptions of the measurement errors on estimating ${\beta}_{1\tau}(t)$ with $n=500$ and an AR(1) structure of the covariance functions with $\rho= 0.5$, $\sigma_{X_1}(t)=3$, $\sigma_{U_1}(t)=2.5$, $\beta_2=0.5$, $\sigma_{X_2}=0.5$, and $\sigma_{U_2}=0.25$. The parameters for the distributions of the error-free covariates were $\sigma_{Z_c}=0.5$ and $p_{Z_b}=0.6$. We considered distributions of Normal, \textit{t}, and Laplace for $U_{1}(t)$ and $U_{2}$.}

In the third simulation study, we investigated the impact of different covariance structures for the functional measurement error, $U_{1ij}(t)$, on estimation accuracy when the covariance structures of the true functional covariate, $X_{1i}(t)$, were known. We set the correlations for the error structures of $U_{1ij}(t)$ and $X_{1i}(t)$ to be $\rho= 0.5$ with $n=500$, $\sigma_{X_1}(t)=3$, $U_{1}(t) \sim MVN\{0, \sigma_{U_1}(t)^2\}$ with $\sigma_{U_1}(t)= 2.5$, $\beta_2=0.5$, $\sigma_{X_2}=0.5$, $U_{2} \sim N(0, \sigma_{U_2}^{2})$ with $\sigma_{U_2}=0.25$, and the parameters for the distributions of the error-free covariates were $\sigma_{Z_c}=0.5$ and $p_{Z_b}=0.6$. We ran these simulations separately with CS, squared exponential (SE), AR(1), independent (IND), and UN error structures.

In the fourth simulation, we investigated the impact of the magnitude of the serial correlations for the functional covariates at different time points and the covariance structures of $X_{1i}(t)$ and $W_{1ij}(t)$ on the estimated parameters. We considered correlations of $\rho_{X1}=\rho_{U1} =(0.25, 0.50, 0.75)$ for the covariance matrices for $\varepsilon_{X_1}(t)$ and $U_1(t)$, with $n=500$, $\sigma_{X_1}(t)=3$, $U_{1}(t) \sim MVN\{0, \sigma_{U_1}(t)^2\}$ with $\sigma_{U_1}(t)= 2.5$, $\sigma_{X_2}=0.5$,  $U_{2} \sim N(0, \sigma_{U_2}^{2})$ with $\sigma_{U_2}=0.25$, $\beta_2=0.5$, $\sigma_{Z_c}=0.5$, and $p_{Z_b}=0.6$. We ran these simulations separately with CS, AR(1), and UN error structures.

For the fifth simulation study, we evaluated the effect of the magnitudes of the measurement errors on the estimated functional coefficient. $U_{1}(t)$ and $U_{2}$ were normally distributed. We considered magnitudes of $\sigma_{X_i}=(1.0, 1.5, 2.0, 4.0)$, $\sigma_{U_i}=(0.5, 1.0, 2.0)$, and $Ratio=\sigma_{X_i}/\sigma_{U_i}$, with $n=500$, $\rho =0.5$, $\beta_2=0.5$, $\sigma_{Z_c}=0.5$, $p_{Z_b}=0.6$, and an AR(1) error structure for the covariance matrices for $\varepsilon_{X_1}(t)$ and $U_1(t)$.

For the sixth simulation study, we evaluated the effect of the scalar coefficients on estimating the functional coefficient. We set the structure of the covariance matrices for the functional terms to be AR(1) with $\rho = 0.5$, $\sigma_{X_1}(t) = 3$, $U_{1}(t) \sim MVN\{0, \sigma_{U_1}(t)^2\}$ with $\sigma_{U_1}(t)= 2.5$, $\sigma_{X_2}=0.5$, $U_{2} \sim N(0, \sigma_{U_2}^{2})$ with $\sigma_{U_2}=0.25$, $\sigma_{Z_c}=0.5$, and $p_{Z_b}=0.6$. We considered $\beta_2=(0.5, 1, 1.5, 2, 4)$.

To assess the performance of the estimators in estimating the true $\beta_{1\tau}(t)$ functions at different quantile levels, we compared the performance of the SIMEX, FUI, and FSMI approaches with the oracle and two other estimators. The oracle estimator is based on the true covariate $X_{1i}(t)$. The other estimators include the average $W_{1ij}(t)$ across all replicates or days {average estimator (Ave)} and a surrogate measure from a single day (naive estimator). We compared all estimators to the true $\beta_{1\tau}(t)$ functions across all quantile levels we evaluated in the simulations. We made similar comparisons for the scalar covariate prone to measurement error and the error-free covariates.

For each simulation condition, we performed $R=500$ iterations and then computed the average estimated parameters across all iterations: $\overline{\beta}_{1\tau}(t) = \frac{1}{R} \sum_{r=1}^{R} \beta_{1\tau}^{(r)}(t)$, $\overline{\beta}_{2\tau} = \frac{1}{R} \sum_{r=1}^{R} \beta_{2}^{(r)}(\tau)$, and $\overline{\gamma}_{p}(\tau) = \frac{1}{R} \sum_{r=1}^{R} \gamma_{p}^{(r)}(\tau), p=1,\hdots, P, \text{ and }  r=1,\hdots, R$. To measure the performance of the different estimators in estimating $\beta_{1\tau}(t)$, we calculated the average squared bias ($ABias^{2}$), average sample variance ($Avar$), and the average integrated mean square error ($AIMSE$), which we define as follows for $\left\{t_{l}\right\}_{l}^{n_{grid}}$, a sequence of equally spaced grid points on $[0,1]$:
\begin{eqnarray}
ABias^2\{\widehat{\beta}_{1\tau}(t)\} &=&
                               \frac{1}{n_{grid}} \sum_{l=1}^{n_{grid}} \{\overline{\beta}_{1\tau}(t_{l})
                              -{{{\beta_{{1\tau}}(t_{l})}}}\}^{2}, \\
AVar\{\widehat{\beta}_{1\tau}(t)\} &=&      \frac{1}{R}\sum_{r=1}^{R}\frac{1}{n_{grid}}\sum_{l=1}^{n_{grid}}\left\{\widehat{\beta}_{1\tau}^{(r)}(t_{l})-\overline{{\beta}}_{1\tau}(t_{l})\right\}^{2}, \text{ and } \\
AIMSE\{\widehat{\beta}_{1\tau}(t)\} &=& ABias^2\{\widehat{\beta}_{1\tau}(t)\}+Avar\{\widehat{\beta}_{1\tau}(t)\}.
\end{eqnarray}
We computed parallel performance measures for the scalar coefficient prone to error:
\begin{eqnarray}
 |Bias\{\widehat{\beta}_{2\tau}\}|= |\overline{\beta}_{2\tau}-\beta_{{2\tau,true}}|,\\
 Var\{\widehat{\beta}_{2\tau}\}= \frac{1}{R}\sum_{r=1}^{R}\left\{\widehat{\beta}_{2\tau}^{(r)}-\overline{{\beta}}_{2\tau}\right\}^{2}, \text{ and }\\
 AIMSE\{\widehat{\beta}_{2\tau}\}=Bias\{\widehat{\beta}_{2\tau}\}^{2}+Var\{\widehat{\beta}_{2\tau}\}.
\end{eqnarray}
Small values of the performance measures indicate good performance and large values indicate poor performance.


\subsection{Simulation Results} \label{subsec:simresult}
\subsubsection{Simulation Study 1: Impact of sample size} \label{subsec:simsetting1}
Table~\ref{table:samplesize} summarizes the performance of the estimators in estimating the functional regression coefficients \{$\widehat{\beta}_{1\tau}(t)$\} in response to varying sample size. For all estimators, increasing the sample size showed large reductions in $Avar$. We also observed decreasing $ABias^{2}$'s for all the estimators. The bias reduction of the SIMEX, FUI, FSMI, and Oracle estimators was nearly identical in all conditions and superior to that of the Ave estimator, which had a much lower bias than the naive estimator. In general, the $Avar$ for the oracle estimator was the smallest followed in ascending order by the Ave, FSMI, FUI, naive, and SIMEX estimators across all quantiles. While the biases associated with the error-corrected methods tended to be smaller than the Ave and naive estimators, the $Avar$ for the Ave estimator was consistently smaller than $Avar$ for the error-corrected methods considered. Additionally, the $Avar$ for the naive estimator was consistently smaller than the $Avar$ of the SIMEX estimators across all quantiles. $AIMSE$ was lowest for the Oracle estimator, followed in ascending order by the Ave, FSMI, FUI, SIMEX, and naive estimators across all quantiles. In summary, we observed a bias-variance trade-off with the SIMEX, FUI, and FSMI when estimating \{$\widehat{\beta}_{1\tau}(t)$\}.

Table~\ref{table:samplesize_W2} shows the performance of the estimators in estimating the scalar coefficients \{$\widehat{\beta}_{2\tau}$\} for the $50$th quantile. The $|Bias|$ for the oracle, SIMEX, FUI, and FSMI estimators all approach zero as the sample size increased.  FUI and FSMI methods had the largest $|Bias|$ when $n = 100$, while the $|Bias|$ for the SIMEX estimator was larger when $n = 500$. We did not observe an improvement in $|Bias|$ with increasing sample size for the Ave and naive estimators. At each sample size, the Ave and naive estimators had moderate and relatively larger $|Bias|$ than the other four estimators, respectively. As sample size increased, $Var$ and $AIMSE$ decreased for all estimators. The Oracle estimator performed the best on these measures ($Var$ and $AIMSE$), followed by the Ave, FUI, FSMI, and SIMEX estimators. The naive estimator had the highest $Var$ and $AIMSE$. 

\begin{table}[h]
\caption{The impact of varying sample sizes on the performance of six estimators in estimating  $\widehat{\beta}_{1\tau}(t)$ in simulation study 1, with an AR(1) error structure, $\rho=0.5$, $\sigma_{X_1}(t)=3$, $U_{1}(t) \sim MVN(0, 2.5^2)$, $\sigma_{X_2}=0.5$, $U_{2} \sim N(0, 0.25^{2})$, $\beta_2=0.5$, $\sigma_{Z_c}=0.5$, and $p_{Z_b}=0.6$.}
\label{table:samplesize}
\centering
\resizebox{\textwidth}{!}{
\begin{tabular}{r|cccccc|cccccc|cccccc}
\hline
\multicolumn{19}{c}{25th Quantile}\\
\hline
&\multicolumn{6}{c|}{$ABias^{2}$}&\multicolumn{6}{c|}{$Avar$}&\multicolumn{6}{c}{$AIMSE$}\\
\hline
n&Oracle&SIMEX&FUI&FSMI&Ave&Naive&Oracle&SIMEX&FUI&FSMI&Ave&Naive&Oracle&SIMEX&FUI&FSMI&Ave&Naive\\
\hline
100&.0044&.0044&.0043&.0045&.0084&.0874&.0048&.0201&.0119&.0115&.0098&.0190&.0092&.0245&.0162&.0160&.0182&.1064\\
500&.0043&.0043&.0043&.0044&.0083&.0861&.0011&.0036&.0023&.0023&.0019&.0037&.0054&.0080&.0066&.0066&.0102&.0898\\
1000&.0040&.0039&.0039&.0040&.0079&.0866&.0008&.0022&.0015&.0014&.0013&.0021&.0048&.0061&.0055&.0055&.0092&.0887\\
2000&.0031&.0031&.0031&.0032&.0070&.0855&.0010&.0017&.0013&.0013&.0011&.0012&.0041&.0048&.0044&.0045&.0081&.0866\\
5000&.0009&.0009&.0009&.0010&.0051&.0843&.0012&.0015&.0014&.0013&.0011&.0008&.0021&.0024&.0023&.0024&.0062&.0851\\
\hline
\multicolumn{19}{c}{50th Quantile}\\
\hline
&\multicolumn{6}{c|}{$ABias^{2}$}&\multicolumn{6}{c|}{$Avar$}&\multicolumn{6}{c}{$AIMSE$}\\
\hline
n&Oracle&SIMEX&FUI&FSMI&Ave&Naive&Oracle&SIMEX&FUI&FSMI&Ave&Naive&Oracle&SIMEX&FUI&FSMI&Ave&Naive\\
\hline
100&.0045&.0045&.0044&.0045&.0085&.0884&.0041&.0164&.0102&.0097&.0084&.0164&.0086&.0208&.0146&.0142&.0169&.1048\\
500&.0045&.0044&.0044&.0045&.0083&.0868&.0009&.0031&.0020&.0019&.0017&.0029&.0054&.0075&.0065&.0064&.0100&.0897\\
1000&.0040&.0040&.0040&.0041&.0081&.0866&.0007&.0019&.0014&.0013&.0011&.0017&.0048&.0059&.0053&.0054&.0092&.0884\\
2000&.0032&.0032&.0032&.0033&.0072&.0855&.0009&.0014&.0012&.0012&.0010&.0010&.0041&.0046&.0044&.0044&.0082&.0866\\
5000&.0008&.0008&.0008&.0010&.0049&.0842&.0012&.0014&.0013&.0013&.0011&.0007&.0020&.0023&.0022&.0022&.0060&.0849\\
\hline
\multicolumn{19}{c}{75th Quantile}\\
\hline
&\multicolumn{6}{c|}{$ABias^{2}$}&\multicolumn{6}{c|}{$Avar$}&\multicolumn{6}{c}{$AIMSE$}\\
\hline
n&Oracle&SIMEX&FUI&FSMI&Ave&Naive&Oracle&SIMEX&FUI&FSMI&Ave&Naive&Oracle&SIMEX&FUI&FSMI&Ave&Naive\\
\hline
100&.0043&.0045&.0044&.0045&.0085&.0869&.0047&.0196&.0119&.0114&.0098&.0190&.0091&.0241&.0162&.0159&.0183&.1058\\
500&.0043&.0043&.0042&.0043&.0082&.0854&.0011&.0037&.0024&.0023&.0020&.0038&.0054&.0079&.0066&.0066&.0101&.0892\\
1000&.0038&.0037&.0037&.0038&.0077&.0870&.0009&.0023&.0016&.0016&.0014&.0019&.0047&.0060&.0053&.0053&.0090&.0890\\
2000&.0029&.0028&.0028&.0030&.0069&.0853&.0011&.0018&.0015&.0014&.0012&.0012&.0040&.0047&.0043&.0044&.0081&.0865\\
5000&.0006&.0006&.0006&.0007&.0047&.0837&.0011&.0014&.0013&.0012&.0011&.0008&.0017&.0020&.0019&.0019&.0057&.0844\\
\hline
\multicolumn{19}{c}{95th Quantile}\\
\hline
&\multicolumn{6}{c|}{$ABias^{2}$}&\multicolumn{6}{c|}{$Avar$}&\multicolumn{6}{c}{$AIMSE$}\\
\hline
n&Oracle&SIMEX&FUI&FSMI&Ave&Naive&Oracle&SIMEX&FUI&FSMI&Ave&Naive&Oracle&SIMEX&FUI&FSMI&Ave&Naive\\
\hline
100&.0030&.0032&.0030&.0031&.0073&.0895&.0123&.0638&.0305&.0288&.0252&.0461&.0153&.0669&.0336&.0319&.0326&.1356\\
500&.0030&.0029&.0029&.0030&.0068&.0835&.0029&.0121&.0065&.0063&.0054&.0090&.0059&.0150&.0094&.0092&.0122&.0925\\
1000&.0026&.0026&.0026&.0027&.0067&.0855&.0020&.0063&.0038&.0036&.0031&.0048&.0046&.0089&.0064&.0064&.0098&.0904\\
2000&.0019&.0020&.0019&.0021&.0061&.0855&.0017&.0037&.0025&.0023&.0020&.0028&.0036&.0057&.0044&.0044&.0081&.0882\\
5000&.0006&.0006&.0006&.0007&.0047&.0837&.0013&.0021&.0016&.0015&.0013&.0013&.0018&.0027&.0022&.0023&.0061&.0850\\
\hline
\end{tabular}}
\end{table}

\begin{table}[h]
\caption{The impact of varying sample sizes on the performance of six estimators in estimating $\widehat{\beta}_{2\tau}$ in simulation study 1, with an AR(1) error structure, $\rho=0.5$, $\sigma_{X_1}(t)=3$, $U_{1}(t) \sim MVN(0, 2.5^2)$, $\sigma_{X_2}=0.5$, $U_{2} \sim N(0, 0.25^{2})$, $\beta_2=0.5$, $\sigma_{Z_c}=0.5$, and $p_{Z_b}=0.6$.}
\label{table:samplesize_W2}
\centering
\resizebox{\textwidth}{!}{
\begin{tabular}{r|cccccc|cccccc|cccccc}
\hline
\multicolumn{19}{c}{25th Quantile}\\
\hline
&\multicolumn{6}{c|}{$|Bias|$}&\multicolumn{6}{c|}{$Var$}&\multicolumn{6}{c}{$AIMSE$}\\
\hline
n&Oracle&SIMEX&FUI&FSMI&Ave&Naive&Oracle&SIMEX&FUI&FSMI&Ave&Naive&Oracle&SIMEX&FUI&FSMI&Ave&Naive\\
\hline
100&.0004&.0004&.0024&.0023&.0152&.1008&.0010&.0041&.0023&.0023&.0021&.0052&.0010&.0041&.0023&.0023&.0023&.0153\\
500&.0003&.0007&.0003&.0002&.0177&.0986&.0002&.0007&.0004&.0004&.0004&.0010&.0002&.0007&.0004&.0004&.0007&.0107\\
1000&.0002&.0001&.0001&.0000&.0174&.1006&.0001&.0003&.0002&.0002&.0002&.0005&.0001&.0003&.0002&.0002&.0005&.0106\\
2000&.0001&.0009&.0003&.0003&.0176&.0995&.0000&.0002&.0001&.0001&.0001&.0002&.0000&.0002&.0001&.0001&.0004&.0101\\
5000&.0003&.0009&.0005&.0005&.0178&.1003&.0000&.0001&.0000&.0000&.0000&.0001&.0000&.0001&.0000&.0000&.0004&.0101\\
\hline
\multicolumn{19}{c}{50th Quantile}\\
\hline
&\multicolumn{6}{c|}{$|Bias|$}&\multicolumn{6}{c|}{$Var$}&\multicolumn{6}{c}{$AIMSE$}\\
\hline
n&Oracle&SIMEX&FUI&FSMI&Ave&Naive&Oracle&SIMEX&FUI&FSMI&Ave&Naive&Oracle&SIMEX&FUI&FSMI&Ave&Naive\\
\hline
100&.0014&.0012&.0024&.0016&.0155&.1028&.0008&.0029&.0017&.0018&.0016&.0041&.0008&.0029&.0018&.0018&.0018&.0146\\
500&.0001&.0014&.0007&.0007&.0167&.0993&.0001&.0006&.0004&.0004&.0003&.0009&.0001&.0006&.0004&.0004&.0006&.0107\\
1000&.0001&.0002&.0000&.0001&.0173&.1004&.0001&.0003&.0002&.0002&.0001&.0004&.0001&.0003&.0002&.0002&.0004&.0104\\
2000&.0002&.0003&.0002&.0003&.0171&.0999&.0000&.0001&.0001&.0001&.0001&.0002&.0000&.0001&.0001&.0001&.0004&.0102\\
5000&.0001&.0004&.0003&.0003&.0175&.1006&.0000&.0001&.0000&.0000&.0000&.0001&.0000&.0001&.0000&.0000&.0003&.0102\\
\hline
\multicolumn{19}{c}{75th Quantile}\\
\hline
&\multicolumn{6}{c|}{$|Bias|$}&\multicolumn{6}{c|}{$Var$}&\multicolumn{6}{c}{$AIMSE$}\\
\hline
n&Oracle&SIMEX&FUI&FSMI&Ave&Naive&Oracle&SIMEX&FUI&FSMI&Ave&Naive&Oracle&SIMEX&FUI&FSMI&Ave&Naive\\
\hline
100&.0001&.0054&.0042&.0032&.0138&.1049&.0009&.0036&.0020&.0020&.0019&.0048&.0009&.0036&.0020&.0020&.0021&.0158\\
500&.0002&.0011&.0002&.0001&.0175&.0999&.0002&.0007&.0004&.0004&.0004&.0010&.0002&.0007&.0004&.0004&.0007&.0109\\
1000&.0005&.0003&.0003&.0001&.0176&.0999&.0001&.0004&.0002&.0002&.0002&.0005&.0001&.0004&.0002&.0002&.0005&.0104\\
2000&.0005&.0002&.0004&.0002&.0176&.1003&.0000&.0002&.0001&.0001&.0001&.0002&.0000&.0002&.0001&.0001&.0004&.0103\\
5000&.0003&.0007&.0004&.0004&.0177&.1014&.0000&.0001&.0000&.0000&.0000&.0001&.0000&.0001&.0000&.0000&.0003&.0104\\
\hline
\multicolumn{19}{c}{95th Quantile}\\
\hline
&\multicolumn{6}{c|}{$|Bias|$}&\multicolumn{6}{c|}{$Var$}&\multicolumn{6}{c}{$AIMSE$}\\
\hline
n&Oracle&SIMEX&FUI&FSMI&Ave&Naive&Oracle&SIMEX&FUI&FSMI&Ave&Naive&Oracle&SIMEX&FUI&FSMI&Ave&Naive\\
\hline
100&.0035&.0056&.0048&.0045&.0136&.0977&.0021&.0101&.0048&.0047&.0044&.0118&.0021&.0101&.0048&.0048&.0046&.0213\\
500&.0002&.0002&.0003&.0002&.0174&.1011&.0004&.0020&.0010&.0010&.0009&.0021&.0004&.0020&.0010&.0010&.0012&.0124\\
1000&.0022&.0042&.0019&.0019&.0190&.1001&.0002&.0009&.0004&.0004&.0004&.0011&.0002&.0009&.0004&.0004&.0008&.0112\\
2000&.0009&.0008&.0006&.0006&.0178&.0996&.0001&.0005&.0002&.0002&.0002&.0005&.0001&.0005&.0002&.0002&.0005&.0104\\
5000&.0003&.0000&.0000&.0000&.0173&.0994&.0000&.0002&.0001&.0001&.0001&.0002&.0000&.0002&.0001&.0001&.0004&.0101\\
\hline
\end{tabular}}
\end{table}

\textcolor{black}{\subsubsection{Simulation Study 2: Impact of measurement error distributions} \label{subsec:simsetting2}}
\textcolor{black}{Table~\ref{table:me_dist} shows the performance of the estimators in estimating \{$\widehat{\beta}_{1\tau}(t)$\} at the $50$th quantile with $n=500$ for different measurement error distributions. For all estimators, the values of the $ABias^{2}$ were similar across the different distributions assumed for the measurements. However, we observed the smallest $Avar$ under the normality assumptions and the largest $Avar$ with the Laplace distributions. The Oracle, SIMEX, FUI, and FSMI estimators all performed better than the Ave and naive estimators when comparing their $ABias^{2}$ values. However, the Ave estimator method had the smallest $Avar$. Thus, all the measurement error corrected methods were robust to the assumed measurement error distributions for the functional covariate.} 

\textcolor{black}{Table~\ref{table:me_dist_W2} shows the performance of the estimators in estimating $\widehat{\beta}_{2\tau}$ under varying distributional assumptions for the measurement errors. We observed the lowest $|Bias|$'s for the estimators under the normality assumption in the following ascending order: Oracle, FUI/FSMI, SIMEX, Ave, and naive. The largest $|Bias|$'s were observed under the Laplace distributions for the FUI, FSMI, SIMEX, Ave, and naive estimators. While the $|Bias|$'s for the measurement error correction approaches were smaller than the Ave and the naive estimators, the $Var$ for the Ave and naive methods were smaller than those of the measurement error correction methods, illustrating a bias-variance tradeoff. The SIMEX-based estimation of the scalar coefficient associated with the error-prone scalar covariate was more sensitive to the assumed distributions for the measurement errors when compared to the Oracle, FUI, and FSMI approaches. While the $ABias^{2}$ associated with the SIMEX-based approach was equivalent to the FUI and FSMI methods for the functional covariate, its $|Bias|$ for the scalar covariate under the different distributional assumptions were larger when compared to those estimated under the FUI and FSMI methods. However, the SIMEX-based estimator had smaller $Var$ values under the \textit{t} and Laplace distributions. In summary, the error-prone scalar covariate is more sensitive to the distributional assumptions of the measurement errors when compared to the error-prone functional covariate. }

\begin{table}[h]
\caption{The effect of different measurement error distributions on the performance of six estimators in estimating $\widehat{\beta}_{1\tau}(t)$ at the $50$th quantile in simulation study 2, with $n=500$, $\rho=0.5$, $\sigma_{X_1}(t)=3$, $\sigma_{U_1}(t)=2.5$, $\sigma_{X_2}=0.5$, $\sigma_{U_2}=0.25$, $\beta_2=0.5$, $\sigma_{Z_c}=0.5$, and $p_{Z_b}=0.6$.}
\label{table:me_dist}
\centering
\resizebox{\textwidth}{!}{
\begin{tabular}{r|cccccc|cccccc|cccccc}
\hline
\multicolumn{19}{c}{50th Quantile}\\
\hline
&\multicolumn{6}{c|}{$ABias^{2}$}&\multicolumn{6}{c|}{$Avar$}&\multicolumn{6}{c}{$AIMSE$}\\
\hline
Distribution of $U_1(t)$/$U_{2}$&Oracle&SIMEX&FUI&FSMI&Ave&Naive&Oracle&SIMEX&FUI&FSMI&Ave&Naive&Oracle&SIMEX&FUI&FSMI&Ave&Naive\\
\hline
Normal&.0045&.0044&.0044&.0045&.0083&.0868&.0009&.0031&.0020&.0019&.0017&.0029&.0054&.0075&.0065&.0064&.0100&.0897\\
\textit{t} dist&.0043&.0043&.0043&.0043&.0080&.0866&.0010&.0057&.0036&.0035&.0030&.0045&.0053&.0100&.0079&.0078&.0110&.0911\\
Laplace&.0044&.0043&.0043&.0044&.0080&.0753&.0010&.0063&.0040&.0038&.0033&.0048&.0053&.0106&.0083&.0082&.0113&.0801\\
\hline
\end{tabular}}
\end{table}

\begin{table}[h]
\caption{The effect of different measurement error distributions on the performance of six estimators in estimating $\widehat{\beta}_{2\tau}$ at the $50$th quantile in simulation study 2, with $n=500$, $\rho=0.5$, $\sigma_{X_1}(t)=3$, $\sigma_{U_1}(t)=2.5$, $\sigma_{X_2}=0.5$, $\sigma_{U_2}=0.25$, $\beta_2=0.5$, $\sigma_{Z_c}=0.5$, and $p_{Z_b}=0.6$.}
\label{table:me_dist_W2}
\centering
\resizebox{\textwidth}{!}{
\begin{tabular}{r|cccccc|cccccc|cccccc}
\hline
\multicolumn{19}{c}{50th Quantile}\\
\hline
&\multicolumn{6}{c|}{$|Bias|$}&\multicolumn{6}{c|}{$Var$}&\multicolumn{6}{c}{$AIMSE$}\\
\hline
Distribution of $U_1(t)$/$U_{2}$&Oracle&SIMEX&FUI&FSMI&Ave&Naive&Oracle&SIMEX&FUI&FSMI&Ave&Naive&Oracle&SIMEX&FUI&FSMI&Ave&Naive\\
\hline
Normal&.0001&.0014&.0007&.0007&.0167&.0993&.0001&.0006&.0004&.0004&.0003&.0009&.0001&.0006&.0004&.0004&.0006&.0107\\
\textit{t} dist&.0008&.0668&.0024&.0027&.1831&.4008&.0001&.0010&.0014&.0014&.0004&.0003&.0001&.0055&.0014&.0014&.0339&.1609\\
Laplace&.0003&.1452&.0127&.0130&.2643&.4429&.0001&.0008&.0033&.0034&.0003&.0002&.0001&.0219&.0035&.0035&.0702&.1963\\
\hline
\end{tabular}}
\end{table}

\subsubsection{Simulation Study 3: Impact of functional error structures} \label{subsec:simsetting3}
Table~\ref{table:covariancestructure} shows the performance of the estimators in estimating \{$\widehat{\beta}_{1\tau}(t)$\} at the $50$th quantile level in response to different functional error structures. Results for the other quantiles are in the supplemental materials. In practice, the error structure of $\varepsilon_{X_1}(t)$ is unknown. However, these simulations indicate how the error structure would affect estimation if it were known.
Certain combinations of error structures produced relatively good or poor performance in all estimators. \textcolor{black}{For a given error structure for $\varepsilon_{X_1}(t)$, $ABias^2$, $AVar$, and $AIMSE$ were lowest for each estimator when $U_1(t)$ had the same error structure.} However, when $U_1(t)$ was assumed to have a more complex error structure than $\varepsilon_{X_1}(t)$ such as squared exponential error structure, all the estimators performed relatively poorly. The Oracle estimator had relatively lower $ABias^2$, $AVar$, and $AIMSE$ in all conditions. The SIMEX estimator had very low $ABias^2$. The FUI, FSMI, Ave, and naive estimators' $ABias^2$ varied dramatically by error structure conditions, with very low values in some scenarios and moderate to high values in others. The naive estimator had the poorest performance overall. The SIMEX estimator had roughly higher $AVar$ than the other five estimators. Among these six estimators across error structure conditions, $AIMSE$ was broadly similar for the FUI and FSMI estimators, and the naive estimator always had the highest value. Thus, we observed that the $ABias^2$ for the SIMEX estimator was more robust to the assumed error structure for $U_1(t)$, however, there was a bias-variance trade-off.

\begin{table}[h]
\caption{The effect of different error structures for the functional terms on the performance of six estimators in estimating $\widehat{\beta}_{1\tau}(t)$ at the $50$th quantile in simulation study 3, with $n=500$, $\rho=0.5$, $\sigma_{X_1}(t)=3$, $U_{1}(t) \sim MVN(0, 2.5^2)$, $\sigma_{X_2}=0.5$, $U_{2} \sim N(0, 0.25^{2})$, $\beta_2=0.5$, $\sigma_{Z_c}=0.5$, and $p_{Z_b}=0.6$. The error structures include compound symmetry (CS), squared exponential (SE), AR(1), independent (IND), and unstructured (UN).}
\label{table:covariancestructure}
\centering
\resizebox{\textwidth}{!}{
\begin{tabular}{rr|cccccc|cccccc|cccccc}
\hline
\multicolumn{20}{c}{50th Quantile}\\
\hline
&&\multicolumn{6}{c|}{$ABias^{2}$}&\multicolumn{6}{c|}{$Avar$}&\multicolumn{6}{c}{$AIMSE$}\\
\hline
Structure of $\varepsilon_{X_1}(t)$&Structure of $U_1(t)$&Oracle&SIMEX&FUI&FSMI&Ave&Naive&Oracle&SIMEX&FUI&FSMI&Ave&Naive&Oracle&SIMEX&FUI&FSMI&Ave&Naive\\
\hline
CS&CS&.0047&.0047&.0047&.0050&.0087&.0858&.0029&.0071&.0046&.0044&.0038&.0053&.0076&.0118&.0093&.0093&.0125&.0911\\
CS&SE&.0047&.1316&.2124&.2173&.2320&.4294&.0029&.0095&.0048&.0046&.0040&.0020&.0076&.1411&.2172&.2219&.2360&.4314\\
CS&AR1&.0047&.0147&.0522&.0573&.0731&.3215&.0029&.0096&.0047&.0044&.0039&.0024&.0076&.0242&.0569&.0617&.0770&.3238\\
CS&IND&.0047&.0050&.0083&.0098&.0185&.1692&.0029&.0077&.0046&.0044&.0038&.0044&.0076&.0127&.0129&.0142&.0224&.1736\\
CS&UN&.0046&.0048&.0061&.0086&.0175&.1635&.0030&.0082&.0055&.0050&.0043&.0053&.0076&.0130&.0116&.0136&.0218&.1688\\
SE&CS&.0050&.0050&.0105&.0102&.0048&.0050&.0006&.0014&.0011&.0011&.0009&.0017&.0056&.0065&.0115&.0113&.0057&.0067\\
SE&SE&.0050&.0050&.0051&.0051&.0092&.1006&.0006&.0053&.0032&.0032&.0026&.0064&.0057&.0103&.0082&.0083&.0119&.1070\\
SE&AR1&.0049&.0047&.0070&.0068&.0047&.0246&.0007&.0025&.0017&.0017&.0014&.0029&.0056&.0072&.0087&.0085&.0061&.0275\\
SE&IND&.0050&.0050&.0095&.0102&.0047&.0074&.0006&.0016&.0012&.0012&.0010&.0020&.0056&.0066&.0107&.0114&.0057&.0094\\
SE&UN&.0048&.0049&.0139&.0116&.0046&.0067&.0007&.0017&.0014&.0013&.0010&.0027&.0056&.0065&.0152&.0130&.0057&.0093\\
AR1&CS&.0044&.0044&.0077&.0066&.0045&.0097&.0009&.0017&.0013&.0012&.0011&.0020&.0054&.0061&.0090&.0078&.0056&.0117\\
AR1&SE&.0045&.0078&.0272&.0290&.0431&.2523&.0009&.0057&.0032&.0031&.0026&.0033&.0054&.0135&.0304&.0320&.0457&.2556\\
AR1&AR1&.0045&.0044&.0044&.0045&.0083&.0868&.0009&.0031&.0020&.0019&.0017&.0029&.0054&.0075&.0065&.0064&.0100&.0897\\
AR1&IND&.0045&.0045&.0065&.0061&.0050&.0221&.0009&.0021&.0015&.0015&.0013&.0025&.0054&.0066&.0080&.0075&.0062&.0246\\
AR1&UN&.0044&.0044&.0094&.0068&.0049&.0216&.0009&.0020&.0015&.0014&.0012&.0025&.0054&.0063&.0109&.0082&.0061&.0240\\
IND&CS&.0046&.0046&.0058&.0145&.0057&.0362&.0020&.0040&.0027&.0032&.0023&.0040&.0066&.0086&.0085&.0177&.0079&.0402\\
IND&SE&.0047&.0512&.1183&.0979&.1402&.3777&.0020&.0087&.0045&.0054&.0037&.0024&.0067&.0598&.1228&.1033&.1440&.3802\\
IND&UN&.0045&.0045&.0053&.0125&.0083&.0827&.0021&.0047&.0033&.0038&.0025&.0039&.0066&.0092&.0086&.0163&.0108&.0866\\
UN&CS&.0051&.0051&.0050&.0052&.0074&.0460&.0021&.0051&.0033&.0034&.0029&.0046&.0072&.0102&.0083&.0086&.0103&.0505\\
UN&SE&.0051&.0583&.1344&.1359&.1509&.3886&.0021&.0099&.0051&.0051&.0044&.0029&.0072&.0682&.1395&.1410&.1554&.3915\\
UN&AR1&.0052&.0083&.0265&.0261&.0385&.2386&.0020&.0072&.0035&.0036&.0031&.0028&.0072&.0154&.0300&.0297&.0416&.2414\\
UN&IND&.0052&.0053&.0066&.0064&.0118&.0990&.0020&.0054&.0032&.0032&.0027&.0038&.0072&.0107&.0098&.0096&.0146&.1028\\
UN&UN&.0052&.0055&.0071&.0076&.0128&.0992&.0020&.0054&.0034&.0034&.0028&.0043&.0072&.0108&.0105&.0110&.0156&.1035\\
\hline
\end{tabular}}
\end{table}

\subsubsection{Simulation Study 4: Impact of serially correlated errors} \label{subsec:simsetting4}
Table~\ref{table: correlation} shows the performance of the estimators in estimating \{$\widehat{\beta}_{1\tau}(t)$\} at the $50$th quantile in response to varying serial correlation of the functional errors. Across all estimators, under the UN error structures, as the serial correlation increased, $ABias^{2}$ modestly increased. The FUI estimator had notably higher $ABias^{2}$ for the UN error structures. FSMI's $ABias^{2}$ profile mirrored that of FUI but was slightly higher. The $ABias^{2}$ profile for the Ave estimator was still higher than the Oracle, SIMEX, FUI, and FSMI estimators, while the naive estimator consistently had the highest $ABias^{2}$. For each estimator, as serial correlation increased, $Avar$ rose under CS and UN error structures but decreased under the AR(1) error structure. After the Oracle estimator, the Ave estimator had the lowest $Avar$ across all types of error structures, followed by the FUI, FSMI, SIMEX, and naive estimators. Apart from the Oracle estimator, the FUI and FSMI had the lowest $AIMSE$, the SIMEX and Ave estimators had somewhat higher values, and the naive estimator had the highest $AIMSE$. In summary, the $ABias^{2}$ for the FUI and FSMI methods estimated under the AR(1) and CS were robust to increasing serial correlations when compared to the $ABias^{2}$ from the FUI and FSMI estimators under the UN error structure. However, the $ABias^{2}$ obtained for the SIMEX estimator was robust to increasing serial correlations across all three error structures.

\begin{table}[h]
\caption{The effect of serially correlated error on the performance of six estimators in estimating $\widehat{\beta}_{1\tau}(t)$ at the $50$th quantile in simulation study 4, with $n=500$, $\sigma_{X_1}(t)=3$, $U_{1}(t) \sim MVN(0, 2.5^2)$, $\sigma_{X_2}=0.5$, $U_{2} \sim N(0, 0.25^{2})$, $\beta_2=0.5$, $\sigma_{Z_c}=0.5$, and $p_{Z_b}=0.6$. The error structures include compound symmetry (CS), AR(1), and unstructured.}
\label{table: correlation}
\centering
\resizebox{\textwidth}{!}{
\begin{tabular}{r|cccccc|cccccc|cccccc}
\hline
\multicolumn{19}{c}{$\rho$ = 0.25}\\
\hline
&\multicolumn{6}{c|}{$ABias^{2}$}&\multicolumn{6}{c|}{$Avar$}&\multicolumn{6}{c}{$AIMSE$}\\
\hline
Structure of $\varepsilon_{X_1}(t)/ U_1(t)$&Oracle&SIMEX&FUI&FSMI&Ave&Naive&Oracle&SIMEX&FUI&FSMI&Ave&Naive&Oracle&SIMEX&FUI&FSMI&Ave&Naive\\
\hline
CS&.0046&.0046&.0046&.0048&.0085&.0869&.0021&.0052&.0035&.0036&.0029&.0043&.0068&.0099&.0081&.0084&.0114&.0912\\
AR1&.0045&.0044&.0044&.0050&.0086&.0873&.0013&.0038&.0025&.0026&.0021&.0036&.0058&.0082&.0069&.0076&.0107&.0908\\
Unstructured&.0050&.0051&.0064&.0066&.0119&.0981&.0018&.0048&.0030&.0033&.0025&.0039&.0067&.0099&.0094&.0099&.0144&.1020\\
\hline
\multicolumn{19}{c}{$\rho$ = 0.5}\\
\hline
&\multicolumn{6}{c|}{$ABias^{2}$}&\multicolumn{6}{c|}{$Avar$}&\multicolumn{6}{c}{$AIMSE$}\\
\hline
Structure of $\varepsilon_{X_1}(t)/ U_1(t)$&Oracle&SIMEX&FUI&FSMI&Ave&Naive&Oracle&SIMEX&FUI&FSMI&Ave&Naive&Oracle&SIMEX&FUI&FSMI&Ave&Naive\\
\hline
CS&.0047&.0047&.0047&.0050&.0087&.0858&.0029&.0071&.0046&.0044&.0038&.0053&.0076&.0118&.0093&.0093&.0125&.0911\\
AR1&.0045&.0044&.0044&.0045&.0083&.0868&.0009&.0031&.0020&.0019&.0017&.0029&.0054&.0075&.0065&.0064&.0100&.0897\\
Unstructured&.0052&.0055&.0071&.0076&.0128&.0992&.0020&.0054&.0034&.0034&.0028&.0043&.0072&.0108&.0105&.0110&.0156&.1035\\
\hline
\multicolumn{19}{c}{$\rho$ = 0.75}\\
\hline
&\multicolumn{6}{c|}{$ABias^{2}$}&\multicolumn{6}{c|}{$Avar$}&\multicolumn{6}{c}{$AIMSE$}\\
\hline
Structure of $\varepsilon_{X_1}(t)/ U_1(t)$&Oracle&SIMEX&FUI&FSMI&Ave&Naive&Oracle&SIMEX&FUI&FSMI&Ave&Naive&Oracle&SIMEX&FUI&FSMI&Ave&Naive\\
\hline
CS&.0046&.0047&.0046&.0058&.0087&.0860&.0058&.0129&.0087&.0079&.0072&.0093&.0105&.0176&.0134&.0137&.0160&.0953\\
AR1&.0041&.0041&.0041&.0048&.0076&.0846&.0008&.0029&.0020&.0018&.0017&.0030&.0049&.0070&.0061&.0066&.0093&.0876\\
Unstructured&.0053&.0056&.0076&.0086&.0134&.1013&.0024&.0063&.0040&.0039&.0033&.0045&.0077&.0118&.0116&.0125&.0167&.1059\\
\hline
\end{tabular}}
\end{table}

\subsubsection{Simulation Study 5: Impact of magnitudes of measurement error} \label{subsec:simsetting5}
Table~\ref{table: functional_var_ME} shows the performance of the estimators in estimating $\widehat{\beta}_{1\tau}(t)$ in response to different magnitudes of measurement error for the functional covariate. The $ABias^{2}$ of all the estimators increased as the magnitude of $\sigma_{U_1}(t)$ increased while $\sigma_{X_1}(t)$ was held constant. The naive estimator consistently had the highest $ABias^{2}$. The Oracle estimator had the lowest $Avar$ across conditions followed by the $Avar$ of the Ave estimator. The SIMEX, FUI, and FSMI estimators had lower $ABias^{2}$ than the Ave and naive estimators. 

Table~\ref{table: scalar_var_ME} shows the performance of the estimators in estimating $\widehat{\beta}_{1\tau}(t)$ in response to different magnitudes of the scalar variable's measurement error. For a given estimator, the $ABias^{2}$ did not change meaningfully across the different magnitudes of $\sigma_{X_2}$ and $\sigma_{U_2}$. However, the Oracle, FUI, SIMEX, and FSMI had smaller values of $ABias^{2}$ when compared with the Ave and naive estimators performing more poorly . \textcolor{black}{Apart from the Oracle estimator, the Ave estimator had the smallest $Avar$ across error magnitudes, followed by the FUI, FSMI, SIMEX, and naive estimators in order.} Aside from the Oracle estimator's consistently low $AIMSE$, the FUI and FSMI estimators also had the lower values of  $AIMSE$, followed by the SIMEX and Ave estimators. The naive estimator had the highest $AIMSE$. In summary, we observed that changing the magnitude of $\sigma_{U_2}$ while holding $\sigma_{X_2}$ constant did not impact the $ABias^{2}$ of the estimators, however, the values of their $Avar$ and $AIMSE$ were impacted by this change.

\begin{table}[h]
\caption{The effect of varying magnitudes of the functional variable's measurement error on the performance of the six estimators in estimating $\widehat{\beta}_{1\tau}(t)$ at the $50$th quantile in simulation study 5, with $n=500$, an AR(1) error structure, $\rho=0.5$, $U_{1}(t)$ was normally distributed, $\sigma_{X_2}=0.5$, $U_{2} \sim N(0, 0.25^{2})$, $\beta_2=0.5$, $\sigma_{Z_c}=0.5$, and $p_{Z_b}=0.6$.}
\label{table: functional_var_ME}
\centering
\resizebox{\textwidth}{!}{
\begin{tabular}{rrr|cccccc|cccccc|cccccc}
\hline
\multicolumn{21}{c}{50th Quantile}\\
\hline
&&&\multicolumn{6}{c|}{$ABias^{2}$}&\multicolumn{6}{c|}{$Avar$}&\multicolumn{6}{c}{$AIMSE$}\\
\hline
$\sigma_{X_1}(t)$&$\sigma_{U_1}(t)$&Ratio&Oracle&SIMEX&FUI&FSMI&Ave&Naive&Oracle&SIMEX&FUI&FSMI&Ave&Naive&Oracle&SIMEX&FUI&FSMI&Ave&Naive\\
\hline
1&0.5&2&.0047&.0047&.0047&.0048&.0053&.0245&.0060&.0123&.0079&.0079&.0073&.0118&.0107&.0170&.0125&.0127&.0127&.0363\\
1&1&1&.0047&.0047&.0046&.0052&.0125&.1276&.0060&.0134&.0093&.0084&.0071&.0083&.0107&.0181&.0139&.0136&.0196&.1360\\
1&2&0.5&.0048&.0133&.0048&.0235&.0703&.3184&.0060&.0149&.0150&.0096&.0061&.0038&.0107&.0282&.0198&.0331&.0764&.3223\\
1.5&0.5&3&.0047&.0047&.0047&.0050&.0048&.0097&.0028&.0055&.0035&.0036&.0034&.0062&.0074&.0102&.0082&.0086&.0082&.0158\\
1.5&1&1.5&.0047&.0047&.0046&.0047&.0064&.0515&.0028&.0062&.0040&.0039&.0035&.0058&.0074&.0109&.0086&.0086&.0099&.0573\\
1.5&2&0.75&.0047&.0052&.0047&.0079&.0249&.2050&.0028&.0081&.0060&.0050&.0038&.0040&.0075&.0134&.0107&.0129&.0287&.2090\\
2&0.5&4&.0046&.0046&.0046&.0051&.0047&.0064&.0016&.0033&.0021&.0022&.0021&.0037&.0063&.0079&.0067&.0073&.0067&.0100\\
2&1&2&.0046&.0046&.0046&.0047&.0052&.0245&.0016&.0037&.0023&.0023&.0022&.0040&.0063&.0083&.0069&.0070&.0073&.0284\\
2&2&1&.0047&.0047&.0046&.0052&.0123&.1267&.0016&.0050&.0033&.0030&.0025&.0036&.0063&.0096&.0080&.0082&.0148&.1303\\
4&0.5&8&.0040&.0040&.0040&.0045&.0040&.0041&.0009&.0013&.0010&.0010&.0010&.0014&.0048&.0052&.0049&.0055&.0049&.0055\\
4&1&4&.0041&.0041&.0041&.0046&.0041&.0058&.0008&.0013&.0009&.0010&.0009&.0016&.0049&.0054&.0050&.0056&.0051&.0074\\
4&2&2&.0041&.0041&.0041&.0043&.0047&.0236&.0008&.0017&.0012&.0012&.0011&.0022&.0049&.0059&.0053&.0055&.0058&.0257\\
\hline
\end{tabular}}
\end{table}

\begin{table}[h]
\caption{The effect of varying magnitudes of the scalar variable's measurement error on the performance of the six estimators in estimating $\widehat{\beta}_{1\tau}(t)$ at the $50$th quantile in simulation study 5, with $n=500$, an AR(1) error structure, $\rho=0.5$, $\sigma_{X_1}(t)=3$, $U_{1}(t) \sim MVN(0, 2.5^2)$, $U_{2}$ was normally distributed, $\beta_2=0.5$, $\sigma_{Z_c}=0.5$, and $p_{Z_b}=0.6$.}
\label{table: scalar_var_ME}
\centering
\resizebox{\textwidth}{!}{
\begin{tabular}{rrr|cccccc|cccccc|cccccc}
\hline
\multicolumn{21}{c}{50th Quantile}\\
\hline
&&&\multicolumn{6}{c|}{$ABias^{2}$}&\multicolumn{6}{c|}{$Avar$}&\multicolumn{6}{c}{$AIMSE$}\\
\hline
$\sigma_{X_2}$&$\sigma_{U_2}$&Ratio&Oracle&SIMEX&FUI&FSMI&Ave&Naive&Oracle&SIMEX&FUI&FSMI&Ave&Naive&Oracle&SIMEX&FUI&FSMI&Ave&Naive\\
\hline
1&0.5&2&.0046&.0046&.0046&.0047&.0086&.0867&.0008&.0037&.0024&.0023&.0020&.0044&.0055&.0083&.0070&.0070&.0105&.0911\\
1&1&1&.0046&.0047&.0046&.0047&.0087&.0867&.0008&.0065&.0041&.0039&.0034&.0069&.0055&.0112&.0088&.0086&.0121&.0936\\
1&2&0.5&.0046&.0047&.0046&.0047&.0088&.0868&.0008&.0131&.0082&.0077&.0067&.0094&.0055&.0178&.0128&.0125&.0156&.0962\\
1.5&0.5&3&.0047&.0047&.0047&.0047&.0086&.0864&.0008&.0037&.0023&.0022&.0019&.0046&.0055&.0083&.0070&.0070&.0105&.0910\\
1.5&1&1.5&.0047&.0047&.0047&.0048&.0087&.0862&.0008&.0067&.0043&.0041&.0035&.0086&.0055&.0114&.0089&.0089&.0122&.0948\\
1.5&2&0.75&.0047&.0047&.0047&.0048&.0088&.0869&.0008&.0160&.0097&.0093&.0081&.0149&.0055&.0207&.0144&.0140&.0168&.1018\\
2&0.5&4&.0047&.0047&.0047&.0047&.0086&.0867&.0008&.0037&.0023&.0022&.0019&.0047&.0055&.0083&.0070&.0070&.0105&.0914\\
2&1&2&.0047&.0047&.0047&.0048&.0087&.0865&.0008&.0067&.0043&.0041&.0036&.0095&.0055&.0114&.0090&.0089&.0123&.0960\\
2&2&1&.0047&.0047&.0046&.0048&.0088&.0864&.0008&.0168&.0103&.0099&.0085&.0205&.0055&.0215&.0149&.0146&.0173&.1069\\
4&0.5&8&.0047&.0047&.0047&.0048&.0086&.0866&.0008&.0036&.0023&.0022&.0019&.0048&.0055&.0083&.0070&.0070&.0105&.0914\\
4&1&4&.0047&.0047&.0047&.0048&.0088&.0862&.0008&.0066&.0042&.0040&.0035&.0110&.0055&.0113&.0089&.0089&.0123&.0972\\
4&2&2&.0047&.0047&.0047&.0048&.0089&.0861&.0008&.0181&.0115&.0111&.0095&.0315&.0055&.0228&.0162&.0159&.0184&.1176\\
\hline
\end{tabular}}
\end{table}

\subsubsection{Simulation Study 6: Impact of $\beta_2$} \label{subsec:simsetting6}
Table~\ref{table: beta2} presents the performance of $\widehat{\beta}_{1\tau}(t)$ under different values of $\beta_2$. With $\beta_2$ increasing, $Avar$ increased but $ABias^{2}$ was not impacted by the values of $\beta_2$ for all the estimators. However, the values of $Avar$ and $AIMSE$ for all the estimators were impacted. In general, the Oracle, SIMEX, FUI, and FSMI estimators had similar and smaller $ABias^{2}$ than the Ave and naive estimators, while the Ave method had the smallest $Avar$ across different $\beta_2$.

\begin{table}[h]
\caption{The effect of $\beta_2$ on the performance of $\widehat{\beta}_{1\tau}(t)$ obtained under the six methods: Oracle, SIMEX, FUI, FSMI, Average, and Naive considered for Simulation Study 6. The performance of the estimators for $\beta_2=(0.5, 1.0, 1.5, 2.0, 4.0)$; $\tau=0.5$; with $n=500$, AR(1) error structure, $\rho=0.5$, $\sigma_{X_1}(t)=3$, $U_{1}(t) \sim MVN(0, 2.5^2)$, $\sigma_{X_2}=0.5$, $U_{2} \sim N(0, 0.25^{2})$, $\sigma_{Z_c}=0.5$, and $p_{Z_b}=0.6$.}
\label{table: beta2}
\centering
\resizebox{\textwidth}{!}{
\begin{tabular}{r|cccccc|cccccc|cccccc}
\hline
\multicolumn{19}{c}{50th Quantile}\\
\hline
&\multicolumn{6}{c|}{$ABias^{2}$}&\multicolumn{6}{c|}{$Avar$}&\multicolumn{6}{c}{$AIMSE$}\\
\hline
$\beta_{2}$&Oracle&SIMEX&FUI&FSMI&Ave&Naive&Oracle&SIMEX&FUI&FSMI&Ave&Naive&Oracle&SIMEX&FUI&FSMI&Ave&Naive\\
\hline
0.5&.0045&.0044&.0044&.0045&.0083&.0868&.0009&.0031&.0020&.0019&.0017&.0029&.0054&.0075&.0065&.0064&.0100&.0897\\
1&.0046&.0046&.0046&.0047&.0086&.0867&.0008&.0037&.0024&.0023&.0020&.0044&.0055&.0083&.0070&.0070&.0105&.0911\\
1.5&.0047&.0047&.0047&.0047&.0087&.0860&.0008&.0050&.0031&.0030&.0026&.0065&.0055&.0096&.0078&.0077&.0112&.0925\\
2&.0047&.0047&.0047&.0048&.0087&.0865&.0008&.0067&.0043&.0041&.0036&.0095&.0055&.0114&.0090&.0089&.0123&.0960\\
4&.0047&.0047&.0047&.0048&.0089&.0861&.0008&.0181&.0115&.0111&.0095&.0315&.0055&.0228&.0162&.0159&.0184&.1176\\
\hline
\end{tabular}}
\end{table}

\section{APPLICATION} \label{sec:application}
We applied our methods to data from the 2003-2004 and 2005-2006 cycles of the NHANES to evaluate the association between physical activity intensity and BMI. We aggregated the minute-level physical activity intensity measures into hourly measures and included both weekday and weekend observations. We removed the extreme observations of physical activity intensity observations (those that were more than 3 times the interquartile range above the 3rd quartile). For each participant, we included data from a day only if they were available for the full 24 hours. We included all participants with physical activity data for at least 2 days. \textcolor{black}{Our final analytic sample had 5,064 community-dwelling adults between the ages of 20 to 85 living in the United States, including $53.5\%$ females. Participants' average age was $50.1  (\pm SD=19.1)$ years, and $73.6\%$ were white, $11.7\%$ were black, $11.2\%$ were Hispanic, and $3.4\%$ were of another race/ethnicity. Participants' average BMI was $28.7 (\pm SD=6.5)$ kg/m$^2$.} NHANES monitored participants' daily physical activity intensity with a physical activity monitor (PAM) (Pensacola, Fl.) that detected uniaxial movements and collected and stored the magnitude of acceleration for each minute. Each participant wore the PAM at home for 7 consecutive days to monitor his/her waking activity. However, participants did not wear the PAM during sleep or water-related activities. NHANES collected dietary data in a face-to-face interview and a telephone interview. In these interviews, participants reported their food and liquid intakes in the past 24 hours along with details including times of consumption, forgotten foods, and detail cycle. In the face-to-face interview, NHANES staff guided participants on how to measure dietary intake and gave them measuring cups, spoons, a ruler, and a booklet containing two-dimensional food models. Three to ten days later in a different week and on a different weekday, they conducted the telephone interview.

We included physical activity intensity and dietary fiber intake in our model as functional and scalar covariates prone to measurement error, respectively, and included age, sex, and race as error-free covariates. We applied sample weights to account for the oversampling of racial subgroups per NHANES analytic guidelines \citep{national2013national}. To select the number of terms in the basis expansion $K_n$, we used the BIC for each quantile level of the response when reducing the dimensions of the functional terms in the model. We estimated the model with each of the SIMEX, FUI, FSMI, Ave, and naive estimators as described in Section \ref{sec:simulation}. For the SIMEX method, we used quadratic extrapolation with $\lambda \in (0.0001, 2.0001)$ in increments of $0.05$ for all quantile levels. For each analysis, we used the nonparametric bootstrap with replacement with $500$ replicates to obtain the 95\% confidence intervals for estimated parameters.

\subsection{Application Results} \label{subsec:appresult}
\subsubsection{Association of physical activity with quantile functions of BMI}
\begin{figure}[h]
\centering
\includegraphics[width=7cm,height=8cm]{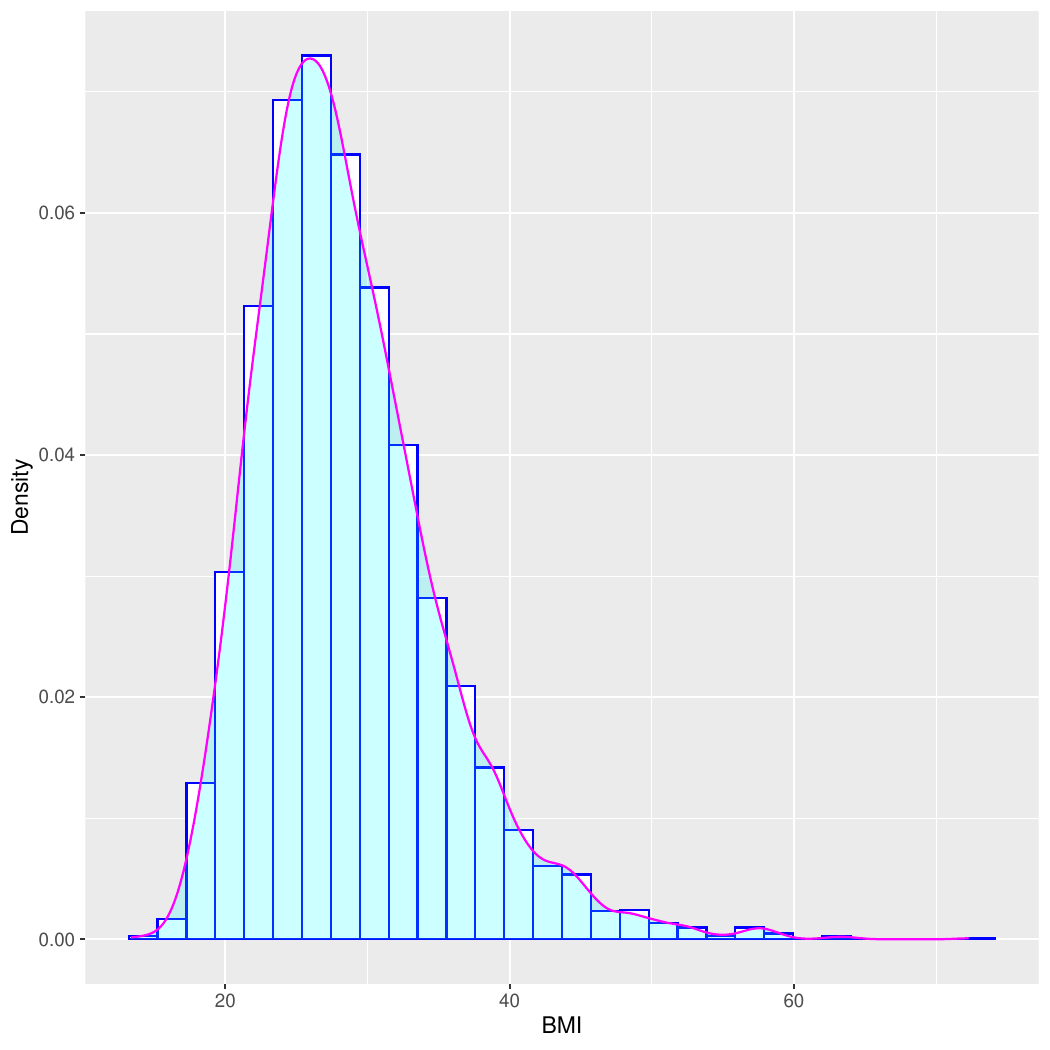}
\caption{Histogram of BMI for the $5,064$ NHANES adult participants.}
\label{bmi_distribution}
\end{figure}

\begin{figure}[h]
\centering
\includegraphics[width=12cm,height=12cm]{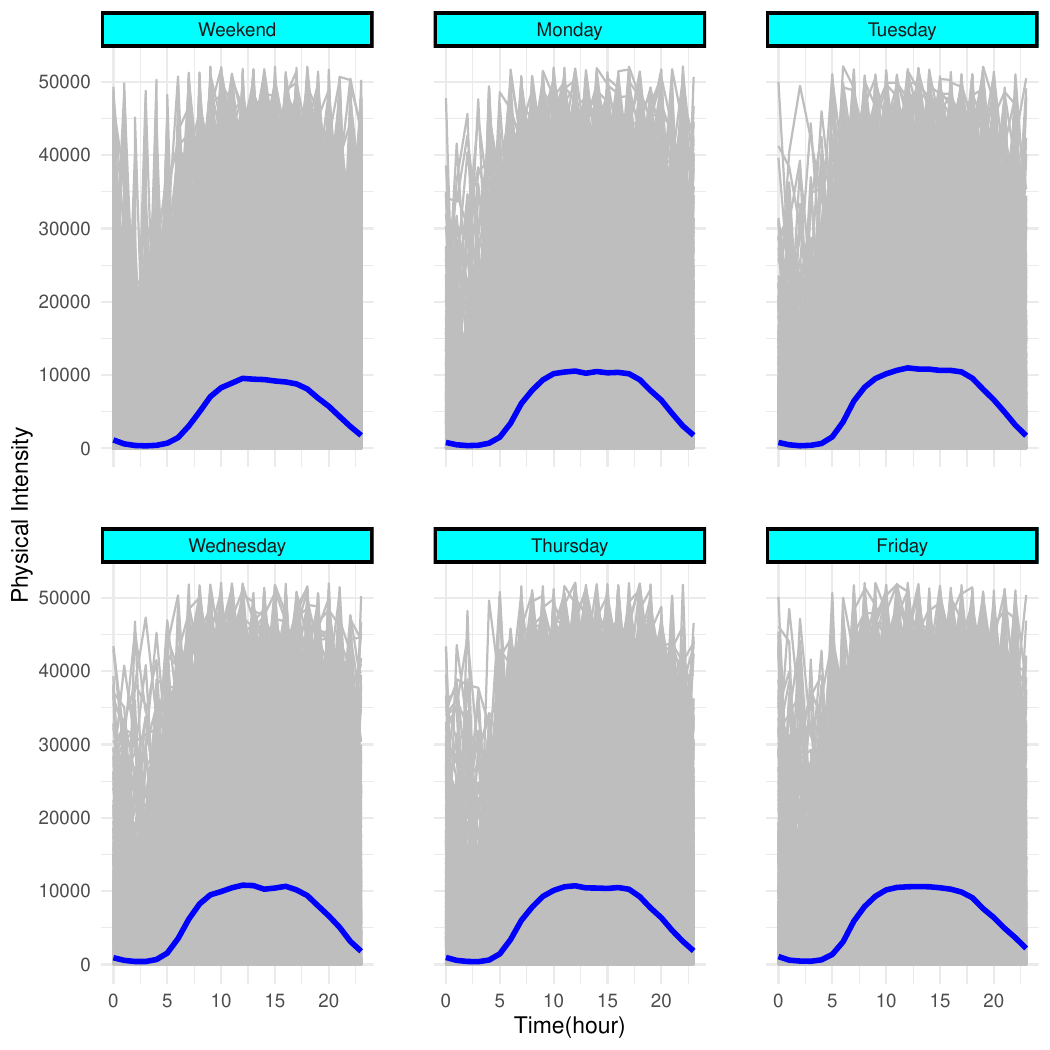}
\caption{Plots of physical activity intensity by day of the week.}
\label{pa_distribution}
\end{figure}

\begin{figure}[h]
\centering
\includegraphics[width=14cm]{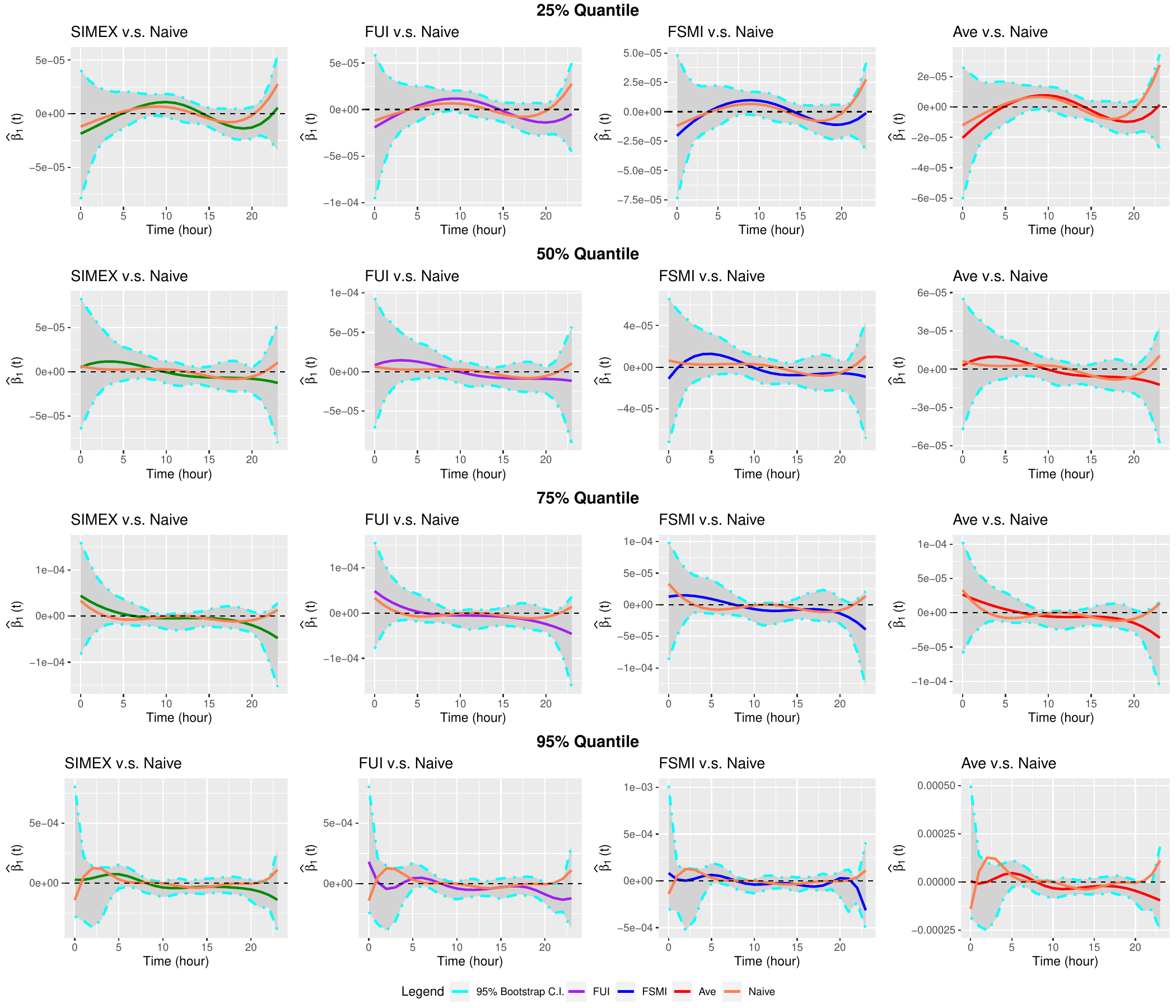}
\caption{Plots of estimated quantile-specific $\beta_{1\tau}(t)$ and 95\% nonparametric bootstrap confidence intervals.}
\label{application_beta_W1}
\end{figure}

Figure~\ref{bmi_distribution} shows that the distribution of the outcome, BMI, in the NHANES data was skewed. The plots in Figure~\ref{pa_distribution} illustrate the nonlinear patterns of physical activity intensity over time and show that the patterns were similar during weekdays and weekends. Figure~\ref{application_beta_W1} shows the quantile-specific estimates of physical activity intensity by time, with comparisons between each of the SIMEX, FUI, FSMI, and Ave estimators with the naive estimator. Portions of the grey-shaded area for confidence intervals that are either completely above or below the dotted black line are statistically significant at the $5\%$ significance level for that time period. The association between physical intensity and BMI depends on the time of the day and quantile levels. We did not observe any statistically significant associations between physical activity intensity and BMI across all the quantiles of BMI considered for all the estimators. However, we observed that the difference between SIMEX, FUI, FSMI, and Ave with the naive estimator depended on the time of day, thus, the influence of measurement error on the estimators is not constant across wear time.  To compare the performance of the SIMEX, FUI, FSMI, and average estimators with the naive estimator, we calculated the quantile-specific average percent difference between the naive estimator and other estimators:
\begin{eqnarray}
\mathrm{Percent\ Difference} &=& mean[|\frac{\beta_{1\tau,\mathrm{with\ ME\ correction}}(t)-\beta_{1\tau,Naive}(t)}{\beta_{1\tau,Naive}(t)}|]*100.
\end{eqnarray}
We compared the other estimators with the naive method because it has the most bias from our simulation studies and it also does not perform any measurement error correction. \textcolor{black}{The SIMEX method performed the best at $25${th} quantile level of BMI with a percent difference of $222\%$; The FUI method performed the best at $50${th} and $95${th} quantile levels of BMI ($221\%$ and $681\%$. respectively); FSMI method performed the best at $75${th} quantile of BMI ($1,913\%$).}


\subsubsection{Association of total fiber intake with quantile functions of BMI}
\begin{figure}[h]
\centering
\includegraphics[width=12cm]{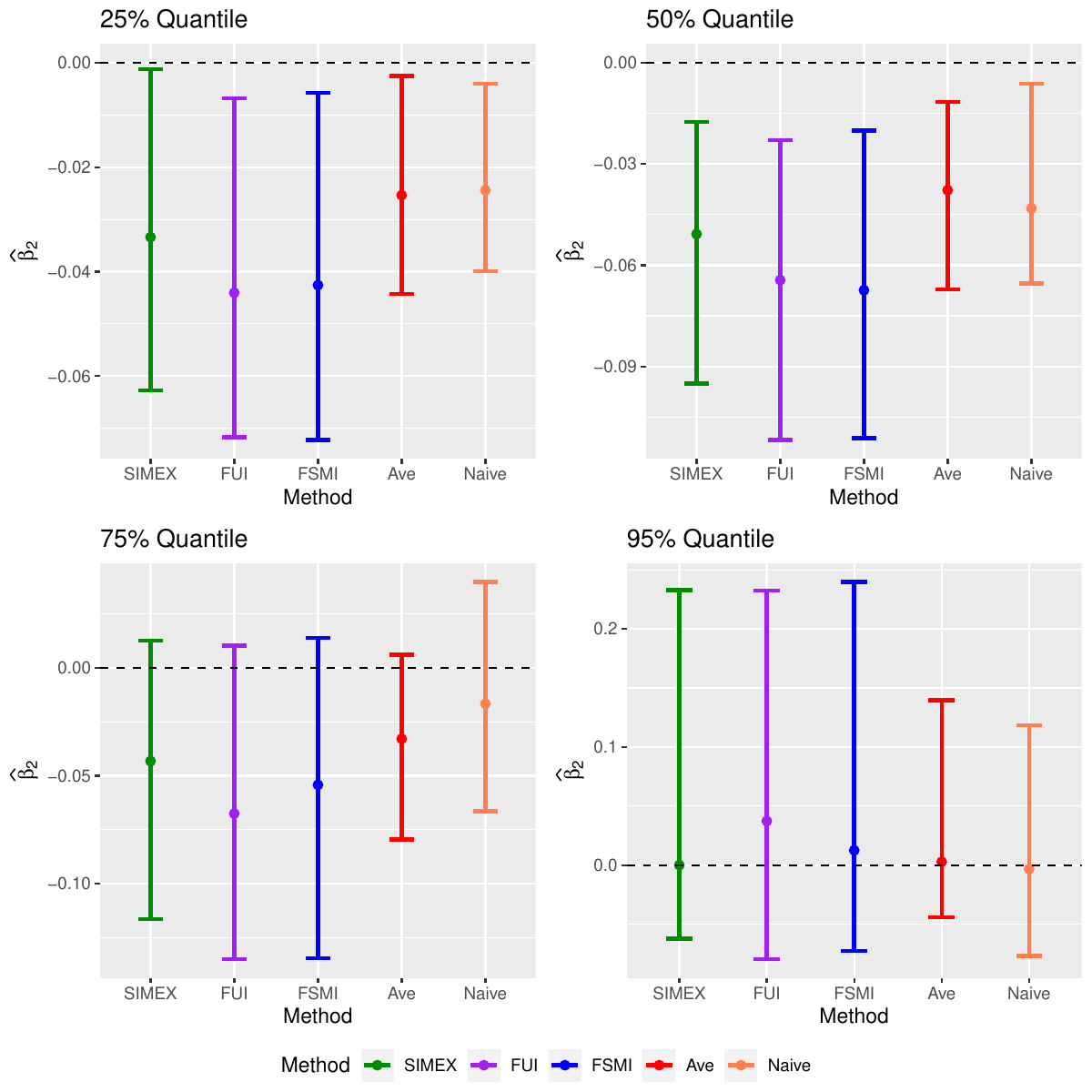}
\caption{Plots of the estimated quantile-specific beta coefficient for fiber intake and 95\% nonparametric bootstrap confidence intervals.}
\label{application_beta_W2}
\end{figure}

Figure~\ref{application_beta_W2} and Table~\ref{table: application_scalar_EF} summarize the associations between dietary fiber intake and the error-free covariates and BMI at different quantile levels. Dietary fiber intake was significantly negatively associated with the $25${th} and $50${th} quantiles of BMI for all estimators. To compare the performance of the estimators for the scalar covariates, we calculated the quantile-specific average percent difference between the naive method and other methods:
\begin{eqnarray}
\mathrm{Percent\ Difference} &=& |\frac{\beta_{2\tau,\mathrm{with\ ME\ correction}}-\beta_{2\tau,Naive}}{\beta_{{2\tau,Naive}}}|*100.
\end{eqnarray}
\textcolor{black}{The FUI method performed the best at the $25${th}, $75${th}, and $95${th} quantile levels of BMI, with percent differences of $80\%$, $304\%$, and $1,255\%$, respectively. The FSMI method performed the best at the $50${th} quantile of BMI, with a percent difference of $56\%$. Overall, we observed that the impact of measurement error on estimating the association between fiber intake and each quantile level considered depended on the quantile. The presence of measurement error attenuated the association between fiber intake and BMI at all the quantile levels considered except for the $95${th} quantile level of BMI. At the $95${th} quantile level of BMI, the FUI and FSMI estimated coefficients were above zero while the SIMEX, Ave, and naive methods coefficients were estimated to be zero at this quantile. }

\textcolor{black}{\subsubsection{Associations of error-free covariates with quantile functions of BMI}
Table~\ref{table: application_scalar_EF} shows the associations between the error-free covariates and BMI for different quantile levels. In general, the association between age and BMI was positive and was statistically significant at the $25${th} and $50${th} quantile levels for all the estimators. Age was not statistically significant with the $75${th} quantile level of BMI for all the estimators, while it was negatively significantly associated with BMI at the $95${th} quantile for the naive estimator only. Blacks had significantly higher BMI than whites at all quantile levels for all estimators, whileHispanics had significantly higher BMI than whites at the $25${th} quantile for all estimators and at the $50${th} quantile for the estimators after measure error correction. For all estimators, women had significantly lower BMI than men at the $25${th} and $50${th} quantiles, but significantly higher BMIs than men at the $95${th} quantile. Women had higher BMI than men at $75${th} quantile under the naive method.}

\begin{table}[ht]
\caption{Estimated associations  of dietary fiber intake and error-free covariates with quantile functions of BMI for the five estimators.}
\label{table: application_scalar_EF}
\centering
\resizebox{\textwidth}{!}{
\begin{tabular}{r *{11}{c}}
\hline
\multicolumn{12}{c}{25th Quantile}\\
\hline
Covariates&&SIMEX&95\% C.I.&FUI&95\% C.I.&FSMI&95\% C.I.&Ave&95\% C.I.&Naive&95\% C.I.\\
\hline
Fiber Intake&&-0.033&(-0.063, -0.001)&-0.044&(-0.072, -0.007)&-0.043&(-0.072, -0.006)&-0.025&(-0.044, -0.003)&-0.024&(-0.040, -0.004)\\
Age&&0.04&(0.03, 0.05)&0.04&(0.03, 0.05)&0.04&(0.03, 0.05)&0.04&(0.03, 0.05)&0.03&(0.02, 0.04)\\
Sex&Male&Ref.\\
&Female&-1.69&(-2.13, -1.24)&-1.69&(-2.07, -1.24)&-1.66&(-2.04, -1.26)&-1.66&(-2.05, -1.26)&-1.42&(-1.92, -1.08)\\
Race&White&Ref.\\
&Hispanic&0.62&(0.14, 1.53)&0.76&(0.18, 1.46)&0.74&(0.17, 1.46)&0.68&(0.16, 1.46)&1.07&(0.29, 1.51)\\
&Black&1.94&(1.21, 2.51)&1.88&(1.30, 2.49)&1.94&(1.30, 2.52)&1.91&(1.28, 2.53)&1.67&(1.08, 2.36)\\
&Other&-1.16&(-2.11, 0.06)&-1.23&(-2.00, -0.12)&-1.19&(-2.03, -0.10)&-1.17&(-2.01, -0.02)&-1.62&(-2.24, -0.42)\\
\hline
\multicolumn{12}{c}{50th Quantile}\\
\hline
Covariates&&SIMEX&95\% C.I.&FUI&95\% C.I.&FSMI&95\% C.I.&Ave&95\% C.I.&Naive&95\% C.I.\\
\hline
Fiber Intake&&-0.051&(-0.095, -0.018)&-0.064&(-0.112, -0.023)&-0.067&(-0.111, -0.020)&-0.038&(-0.067, -0.012)&-0.043&(-0.065, -0.006)\\
Age&&0.04&(0.02, 0.05)&0.04&(0.02, 0.05)&0.04&(0.02, 0.05)&0.04&(0.02, 0.05)&0.03&(0.02, 0.04)\\
Sex&Male&Ref.\\
&Female&-1.16&(-1.73, -0.69)&-1.18&(-1.66, -0.73)&-1.19&(-1.67, -0.71)&-1.15&(-1.67, -0.72)&-1.10&(-1.65, -0.56)\\
Race&White&Ref.\\
&Hispanic&0.91&(0.03, 1.75)&0.93&(0.01, 1.62)&0.95&(0.01, 1.64)&0.94&(0.03, 1.64)&0.70&(-0.14, 1.54)\\
&Black&2.59&(1.72, 3.56)&2.66&(1.83, 3.59)&2.63&(1.83, 3.59)&2.70&(1.80, 3.54)&2.51&(1.67, 3.48)\\
&Other&-0.84&(-2.26, 0.50)&-1.08&(-2.10, 0.43)&-1.02&(-2.11, 0.42)&-0.94&(-2.14, 0.39)&-1.09&(-2.60, 0.31)\\
\hline
\multicolumn{12}{c}{75th Quantile}\\
\hline
Covariates&&SIMEX&95\% C.I.&FUI&95\% C.I.&FSMI&95\% C.I.&Ave&95\% C.I.&Naive&95\% C.I.\\
\hline
Fiber Intake&&-0.043&(-0.116, 0.012)&-0.068&(-0.135, 0.010)&-0.054&(-0.134, 0.014)&-0.033&(-0.079, 0.006)&-0.017&(-0.066, 0.040)\\
Age&&0.00&(-0.02, 0.03)&0.00&(-0.01, 0.03)&0.00&(-0.02, 0.03)&0.00&(-0.02, 0.03)&0.00&(-0.02, 0.02)\\
Sex&Male&Ref.\\
&Female&0.41&(-0.30, 1.18)&0.53&(-0.13, 1.20)&0.48&(-0.17, 1.16)&0.50&(-0.16, 1.14)&0.98&(0.26, 1.84)\\
Race&White&Ref.\\
&Hispanic&-0.09&(-1.17, 0.97)&-0.12&(-1.14, 0.82)&-0.04&(-1.18, 0.82)&-0.05&(-1.17, 0.86)&-0.64&(-1.52, 0.71)\\
&Black&3.17&(1.76, 3.97)&3.04&(1.75, 3.93)&3.09&(1.81, 4.03)&3.11&(1.79, 3.96)&2.67&(1.30, 3.91)\\
&Other&-0.79&(-2.85, 1.05)&-0.54&(-2.57, 1.01)&-0.67&(-2.69, 1.04)&-0.65&(-2.68, 0.98)&-1.19&(-3.29, 1.14)\\
\hline
\multicolumn{12}{c}{95th Quantile}\\
\hline
Covariates&&SIMEX&95\% C.I.&FUI&95\% C.I.&FSMI&95\% C.I.&Ave&95\% C.I.&Naive&95\% C.I.\\
\hline
Fiber Intake&&0.000&(-0.062, 0.233)&0.037&(-0.079, 0.232)&0.013&(-0.073, 0.240)&0.003&(-0.044, 0.140)&-0.003&(-0.077, 0.118)\\
Age&&-0.06&(-0.11, 0.02)&-0.05&(-0.10, 0.02)&-0.05&(-0.10, 0.02)&-0.05&(-0.10, 0.02)&-0.08&(-0.13, -0.01)\\
Sex&Male&Ref.\\
&Female&3.27&(1.12, 4.54)&2.81&(1.38, 4.35)&2.81&(1.47, 4.19)&2.93&(1.34, 4.19)&3.32&(1.99, 5.14)\\
Race&White&Ref.\\
&Hispanic&-1.36&(-3.50, 0.83)&-1.26&(-3.18, 0.56)&-1.15&(-3.24, 0.47)&-1.51&(-3.11, 0.47)&-2.42&(-4.29, 1.44)\\
&Black&2.45&(0.91, 7.13)&3.49&(1.35, 6.71)&3.39&(1.35, 6.63)&3.18&(1.53, 6.74)&2.67&(0.78, 5.70)\\
&Other&-0.23&(-3.47, 2.77)&0.12&(-2.75, 2.19)&-0.22&(-2.84, 2.24)&-0.28&(-2.82, 2.16)&0.28&(-3.02, 2.67)\\
\hline
\end{tabular}}
\end{table}

\section{CONCLUSION}\label{sec:conclusion}
We developed the sparse conditional functional quantile regression model with functional and scalar covariates prone to classical measurement error. We assumed that the measurement errors in the serially observed longitudinal high dimensional functional data have complex heteroscedastic error structures and the scalar covariate prone to error has a scalar measurement error variance. To correct measurement error, we implemented a semiparametric SIMEX approach and two parametric approaches based on mixed effects models. To identify the model, we used repeated measurements on the surrogate measures for the error-prone covariates. Before applying the SIMEX method, we first reduced the dimension of the functional covariate with B-spline expansions. However, the mixed effects-based methods proposed for measurement error did not require any dimension reduction in the measurement error correction step. We investigated the finite sample properties of our proposed methods in simulations. Our simulation results indicated the importance of correcting for measurement error in this setting. \textcolor{black}{The SIMEX, FUI, and FSMI methods performed well in reducing measurement error biases for both the scalar and functional covariates prone to measurement errors, but with some bias-variance tradeoff.} The variances for the FSMI method tended to be smaller than those for the SIMEX and FUI methods. Averaging the error-prone measures across the repeated days of observation also reduced their bias compared to the naive method based on estimation with a single day of data. However, to a lesser degree than the SIMEX, FUI, and FSMI approaches. Our simulations indicated that measurement error correction methods implemented for the functional covariate were more robust to the distributional assumptions of the measurement error, however, the scalar covariate was more sensitive to the assumed distributions for the measurement error. We applied our methods in analyzing the associations of device-measured physical activity intensity and self-reported fiber intake with quantile levels of BMI after adjusting for error-free covariates including age, sex, and race/ethnicity in community-dwelling adults in the U.S. The level of measurement error in the functional covariate depended on the time of day for physical activity and quantile level. We did not observe any statistically significcant associations between device-measured physical activity intensity and the quantile levels of BMI considered across time. There was also no statistically significant association between fiber intake and the quantile levels of BMI.

\clearpage

\noindent {\textbf{Acknowledgement}}\\
The authors would like to thank Dr. Devon Brewer for his constructive feedback and assistance with editing the manuscript. \\

\noindent {\textbf{Funding}}\\
This research was supported by an award from the National Institutes of Diabetes, Digestive, and Kidney Disease Award number R01DK132385. This research was also supported in part by Lilly Endowment, Inc., through its support for the Indiana University Pervasive Technology Institute.\\

\noindent {\textbf{Contribution}}\\
CDT conceived this project, performed some of the data analyses and simulations, drafted the manuscript, and supervised the project. XC performed simulations and data analyses for the project and drafted the manuscript. YL assisted with the data cleaning and simulations. LX and SJ edited the draft and provided feedback on the project. RSZ supervised the simulation studies and data analysis and interpretation.\\

\noindent {\textbf{Disclosure statement}}\\
The authors have declared no conflict of interest.\\

\noindent {\textbf{Availability of Data}}\\ 
The data that support the findings of this study are available from the corresponding author, [CDT], upon reasonable request.\\

\bibliographystyle{tfnlm}
\bibliography{flriv_sim.bib}

\appendix

\section{Additional simulation results for $\widehat{\beta}_{1\tau}(t)$}\label{sec:sampeff}

\input{Supplemental_Materials.tex}

\end{document}

%% file: Supplemental_Materials.tex
\begin{center}
\setlength{\tiny\tabcolsep}{1.2pt} 
\setlength{\LTcapwidth}{\textwidth}
\begin{longtable}{rr|cccccc|cccccc|cccccc}
\caption{The effect of error structures on the performance of $\widehat{\beta}_{1\tau}(t)$ obtained under the six methods: oracle, SIMEX, FUI, FSMI, average, and naive considered for Simulation Study 2. The ABias$^2$, Avar, and AIMSE of the estimators for error structures of compound symmetry (CS), squared exponential (SE), AR(1), independent, and unstructured; $\tau=(0.25,0.5,0.75,0.95)$; with $n=500$, $\rho=0.5$, $\sigma_{X_1}(t)=3$, $\sigma_{X_1}(t)=3$, $U_{1}(t) \sim MVN(0, 2.5^2)$, $\sigma_{X_2}=0.5$, $U_{2} \sim N(0, 0.25^{2})$, $\beta_2=0.5$, $\sigma_{Z_c}=0.5$, and $p_{Z_b}=0.6$.}
\label{table: covariancestructure_full}
\\ \hline
\multicolumn{20}{c}{25th Quantile}\\
\hline
&&\multicolumn{6}{c|}{$ABias^{2}$}&\multicolumn{6}{c|}{$Avar$}&\multicolumn{6}{c}{$AIMSE$}\\
\hline
Structure of $\varepsilon_{X_1}$&Structure of $U_1$&Oracle&SIMEX&FUI&FSMI&Ave&Naive&Oracle&SIMEX&FUI&FSMI&Ave&Naive&Oracle&SIMEX&FUI&FSMI&Ave&Naive\\
\hline
CS&CS&.0046&.0046&.0046&.0049&.0087&.0881&.0036&.0088&.0057&.0053&.0047&.0066&.0083&.0134&.0103&.0102&.0134&.0947\\
CS&SE&.0047&.1308&.2118&.2166&.2315&.4275&.0036&.0119&.0057&.0052&.0047&.0025&.0083&.1428&.2175&.2218&.2363&.4300\\
CS&AR1&.0045&.0134&.0506&.0559&.0714&.3198&.0037&.0110&.0054&.0050&.0044&.0029&.0082&.0245&.0559&.0609&.0758&.3226\\
CS&Independent&.0046&.0048&.0081&.0096&.0183&.1699&.0037&.0097&.0056&.0053&.0047&.0052&.0082&.0145&.0137&.0150&.0230&.1751\\
CS&Unstructured&.0046&.0049&.0062&.0087&.0176&.1635&.0036&.0096&.0062&.0057&.0048&.0061&.0083&.0145&.0124&.0143&.0224&.1696\\
SE&CS&.0048&.0048&.0100&.0098&.0046&.0049&.0009&.0019&.0015&.0015&.0012&.0022&.0057&.0066&.0114&.0112&.0058&.0071\\
SE&SE&.0050&.0050&.0050&.0050&.0090&.1011&.0008&.0070&.0042&.0042&.0035&.0077&.0057&.0120&.0092&.0092&.0125&.1088\\
SE&AR1&.0048&.0047&.0069&.0068&.0047&.0246&.0009&.0031&.0020&.0021&.0017&.0035&.0057&.0078&.0089&.0088&.0063&.0281\\
SE&Independent&.0047&.0047&.0092&.0098&.0044&.0071&.0009&.0021&.0016&.0016&.0013&.0029&.0056&.0068&.0108&.0115&.0057&.0100\\
SE&Unstructured&.0045&.0045&.0133&.0111&.0042&.0062&.0010&.0023&.0020&.0019&.0015&.0034&.0056&.0068&.0152&.0130&.0057&.0096\\
AR1&CS&.0044&.0043&.0076&.0065&.0045&.0097&.0010&.0021&.0015&.0014&.0013&.0023&.0054&.0064&.0091&.0079&.0057&.0120\\
AR1&SE&.0045&.0077&.0271&.0289&.0430&.2527&.0010&.0068&.0038&.0036&.0032&.0040&.0055&.0145&.0309&.0325&.0462&.2567\\
AR1&AR1&.0043&.0043&.0043&.0044&.0083&.0861&.0011&.0036&.0023&.0023&.0019&.0037&.0054&.0080&.0066&.0066&.0102&.0898\\
AR1&Independent&.0043&.0042&.0063&.0058&.0047&.0219&.0011&.0027&.0019&.0019&.0016&.0032&.0054&.0069&.0082&.0077&.0063&.0251\\
AR1&Unstructured&.0041&.0041&.0091&.0064&.0045&.0209&.0012&.0024&.0020&.0018&.0015&.0031&.0053&.0065&.0111&.0082&.0060&.0240\\
Independent&CS&.0044&.0044&.0055&.0141&.0055&.0366&.0025&.0050&.0033&.0039&.0027&.0046&.0069&.0094&.0088&.0180&.0082&.0412\\
Independent&SE&.0047&.0509&.1182&.0975&.1401&.3768&.0023&.0113&.0056&.0067&.0046&.0030&.0070&.0621&.1237&.1041&.1447&.3798\\
Independent&AR1&.0045&.0053&.0155&.0070&.0296&.2254&.0024&.0086&.0046&.0054&.0038&.0036&.0070&.0140&.0201&.0124&.0334&.2290\\
Independent&Independent&.0043&.0044&.0043&.0107&.0083&.0858&.0026&.0064&.0041&.0051&.0034&.0053&.0069&.0108&.0085&.0159&.0118&.0911\\
Independent&Unstructured&.0043&.0044&.0052&.0125&.0081&.0831&.0026&.0056&.0038&.0045&.0030&.0048&.0069&.0100&.0090&.0171&.0111&.0879\\
Unstructured&CS&.0050&.0049&.0049&.0051&.0072&.0460&.0025&.0061&.0037&.0038&.0032&.0055&.0075&.0110&.0087&.0090&.0104&.0515\\
Unstructured&SE&.0050&.0579&.1341&.1353&.1504&.3888&.0024&.0116&.0055&.0057&.0048&.0032&.0074&.0694&.1396&.1409&.1553&.3919\\
Unstructured&AR1&.0050&.0078&.0260&.0255&.0380&.2370&.0025&.0087&.0043&.0043&.0037&.0031&.0075&.0165&.0303&.0298&.0417&.2402\\
Unstructured&Independent&.0050&.0053&.0065&.0063&.0117&.1000&.0025&.0065&.0037&.0039&.0032&.0045&.0075&.0118&.0103&.0102&.0149&.1045\\
Unstructured&Unstructured&.0051&.0054&.0071&.0075&.0127&.0992&.0024&.0068&.0041&.0040&.0034&.0052&.0075&.0122&.0112&.0115&.0161&.1043\\
\hline
\multicolumn{20}{c}{50th Quantile}\\
\hline
&&\multicolumn{6}{c|}{$ABias^{2}$}&\multicolumn{6}{c|}{$Avar$}&\multicolumn{6}{c}{$AIMSE$}\\
\hline
Structure of $\varepsilon_{X_1}$&Structure of $U_1$&Oracle&SIMEX&FUI&FSMI&Ave&Naive&Oracle&SIMEX&FUI&FSMI&Ave&Naive&Oracle&SIMEX&FUI&FSMI&Ave&Naive\\
\hline
CS&CS&.0047&.0047&.0047&.0050&.0087&.0858&.0029&.0071&.0046&.0044&.0038&.0053&.0076&.0118&.0093&.0093&.0125&.0911\\
CS&SE&.0047&.1316&.2124&.2173&.2320&.4294&.0029&.0095&.0048&.0046&.0040&.0020&.0076&.1411&.2172&.2219&.2360&.4314\\
CS&AR1&.0047&.0147&.0522&.0573&.0731&.3215&.0029&.0096&.0047&.0044&.0039&.0024&.0076&.0242&.0569&.0617&.0770&.3238\\
CS&Independent&.0047&.0050&.0083&.0098&.0185&.1692&.0029&.0077&.0046&.0044&.0038&.0044&.0076&.0127&.0129&.0142&.0224&.1736\\
CS&Unstructured&.0046&.0048&.0061&.0086&.0175&.1635&.0030&.0082&.0055&.0050&.0043&.0053&.0076&.0130&.0116&.0136&.0218&.1688\\
SE&CS&.0050&.0050&.0105&.0102&.0048&.0050&.0006&.0014&.0011&.0011&.0009&.0017&.0056&.0065&.0115&.0113&.0057&.0067\\
SE&SE&.0050&.0050&.0051&.0051&.0092&.1006&.0006&.0053&.0032&.0032&.0026&.0064&.0057&.0103&.0082&.0083&.0119&.1070\\
SE&AR1&.0049&.0047&.0070&.0068&.0047&.0246&.0007&.0025&.0017&.0017&.0014&.0029&.0056&.0072&.0087&.0085&.0061&.0275\\
SE&Independent&.0050&.0050&.0095&.0102&.0047&.0074&.0006&.0016&.0012&.0012&.0010&.0020&.0056&.0066&.0107&.0114&.0057&.0094\\
SE&Unstructured&.0048&.0049&.0139&.0116&.0046&.0067&.0007&.0017&.0014&.0013&.0010&.0027&.0056&.0065&.0152&.0130&.0057&.0093\\
AR1&CS&.0044&.0044&.0077&.0066&.0045&.0097&.0009&.0017&.0013&.0012&.0011&.0020&.0054&.0061&.0090&.0078&.0056&.0117\\
AR1&SE&.0045&.0078&.0272&.0290&.0431&.2523&.0009&.0057&.0032&.0031&.0026&.0033&.0054&.0135&.0304&.0320&.0457&.2556\\
AR1&AR1&.0045&.0044&.0044&.0045&.0083&.0868&.0009&.0031&.0020&.0019&.0017&.0029&.0054&.0075&.0065&.0064&.0100&.0897\\
AR1&Independent&.0045&.0045&.0065&.0061&.0050&.0221&.0009&.0021&.0015&.0015&.0013&.0025&.0054&.0066&.0080&.0075&.0062&.0246\\
AR1&Unstructured&.0044&.0044&.0094&.0068&.0049&.0216&.0009&.0020&.0015&.0014&.0012&.0025&.0054&.0063&.0109&.0082&.0061&.0240\\
Independent&CS&.0046&.0046&.0058&.0145&.0057&.0362&.0020&.0040&.0027&.0032&.0023&.0040&.0066&.0086&.0085&.0177&.0079&.0402\\
Independent&SE&.0047&.0512&.1183&.0979&.1402&.3777&.0020&.0087&.0045&.0054&.0037&.0024&.0067&.0598&.1228&.1033&.1440&.3802\\
Independent&AR1&.0047&.0057&.0160&.0074&.0302&.2259&.0020&.0067&.0038&.0044&.0031&.0031&.0067&.0124&.0198&.0118&.0334&.2290\\
Independent&Independent&.0045&.0046&.0046&.0109&.0086&.0864&.0021&.0054&.0035&.0043&.0029&.0044&.0066&.0100&.0081&.0152&.0116&.0908\\
Independent&Unstructured&.0045&.0045&.0053&.0125&.0083&.0827&.0021&.0047&.0033&.0038&.0025&.0039&.0066&.0092&.0086&.0163&.0108&.0866\\
Unstructured&CS&.0051&.0051&.0050&.0052&.0074&.0460&.0021&.0051&.0033&.0034&.0029&.0046&.0072&.0102&.0083&.0086&.0103&.0505\\
Unstructured&SE&.0051&.0583&.1344&.1359&.1509&.3886&.0021&.0099&.0051&.0051&.0044&.0029&.0072&.0682&.1395&.1410&.1554&.3915\\
Unstructured&AR1&.0052&.0083&.0265&.0261&.0385&.2386&.0020&.0072&.0035&.0036&.0031&.0028&.0072&.0154&.0300&.0297&.0416&.2414\\
Unstructured&Independent&.0052&.0053&.0066&.0064&.0118&.0990&.0020&.0054&.0032&.0032&.0027&.0038&.0072&.0107&.0098&.0096&.0146&.1028\\
Unstructured&Unstructured&.0052&.0055&.0071&.0076&.0128&.0992&.0020&.0054&.0034&.0034&.0028&.0043&.0072&.0108&.0105&.0110&.0156&.1035\\
\hline
\multicolumn{20}{c}{75th Quantile}\\
\hline
&&\multicolumn{6}{c|}{$ABias^{2}$}&\multicolumn{6}{c|}{$Avar$}&\multicolumn{6}{c}{$AIMSE$}\\
\hline
Structure of $\varepsilon_{X_1}$&Structure of $U_1$&Oracle&SIMEX&FUI&FSMI&Ave&Naive&Oracle&SIMEX&FUI&FSMI&Ave&Naive&Oracle&SIMEX&FUI&FSMI&Ave&Naive\\
\hline
CS&CS&.0045&.0046&.0045&.0049&.0086&.0863&.0037&.0091&.0059&.0055&.0049&.0064&.0082&.0136&.0104&.0104&.0134&.0927\\
CS&SE&.0047&.1304&.2119&.2163&.2315&.4278&.0035&.0113&.0056&.0054&.0047&.0023&.0082&.1418&.2175&.2217&.2361&.4301\\
CS&AR1&.0046&.0147&.0526&.0576&.0736&.3218&.0036&.0113&.0056&.0053&.0047&.0028&.0082&.0260&.0582&.0629&.0783&.3246\\
CS&Independent&.0045&.0047&.0079&.0094&.0181&.1705&.0037&.0094&.0054&.0052&.0045&.0052&.0082&.0141&.0134&.0146&.0226&.1757\\
CS&Unstructured&.0046&.0048&.0061&.0087&.0177&.1651&.0037&.0097&.0062&.0057&.0048&.0061&.0082&.0145&.0123&.0143&.0225&.1712\\
SE&CS&.0048&.0049&.0102&.0100&.0046&.0049&.0008&.0017&.0013&.0013&.0011&.0022&.0056&.0066&.0115&.0113&.0057&.0070\\
SE&SE&.0050&.0051&.0051&.0051&.0092&.1002&.0007&.0064&.0038&.0038&.0032&.0077&.0057&.0115&.0089&.0089&.0123&.1079\\
SE&AR1&.0049&.0047&.0069&.0068&.0047&.0242&.0008&.0032&.0021&.0021&.0017&.0036&.0057&.0079&.0090&.0089&.0064&.0279\\
SE&Independent&.0048&.0048&.0093&.0100&.0045&.0071&.0008&.0021&.0015&.0015&.0012&.0027&.0057&.0069&.0108&.0116&.0057&.0098\\
SE&Unstructured&.0047&.0047&.0137&.0115&.0044&.0064&.0009&.0021&.0018&.0017&.0013&.0036&.0056&.0068&.0154&.0132&.0057&.0100\\
AR1&CS&.0042&.0042&.0073&.0063&.0043&.0096&.0012&.0021&.0016&.0015&.0013&.0024&.0053&.0063&.0090&.0078&.0056&.0119\\
AR1&SE&.0045&.0077&.0271&.0288&.0430&.2513&.0010&.0067&.0037&.0036&.0031&.0038&.0055&.0144&.0308&.0324&.0461&.2551\\
AR1&AR1&.0043&.0043&.0042&.0043&.0082&.0854&.0011&.0037&.0024&.0023&.0020&.0038&.0054&.0079&.0066&.0066&.0101&.0892\\
AR1&Independent&.0041&.0041&.0061&.0056&.0046&.0215&.0012&.0026&.0020&.0019&.0016&.0033&.0053&.0068&.0081&.0075&.0062&.0249\\
AR1&Unstructured&.0041&.0041&.0092&.0065&.0046&.0216&.0012&.0025&.0020&.0018&.0015&.0029&.0053&.0066&.0112&.0083&.0061&.0245\\
Independent&CS&.0044&.0044&.0056&.0141&.0055&.0373&.0025&.0049&.0033&.0039&.0028&.0045&.0069&.0093&.0089&.0180&.0082&.0418\\
Independent&SE&.0046&.0504&.1176&.0978&.1397&.3771&.0024&.0114&.0056&.0067&.0047&.0029&.0070&.0618&.1233&.1045&.1443&.3800\\
Independent&AR1&.0045&.0055&.0158&.0072&.0301&.2273&.0024&.0086&.0047&.0056&.0039&.0038&.0069&.0140&.0205&.0128&.0340&.2311\\
Independent&Independent&.0044&.0044&.0044&.0106&.0085&.0866&.0025&.0066&.0042&.0052&.0035&.0052&.0069&.0110&.0086&.0158&.0120&.0918\\
Independent&Unstructured&.0043&.0043&.0050&.0121&.0081&.0847&.0025&.0057&.0041&.0049&.0032&.0048&.0068&.0100&.0091&.0170&.0113&.0895\\
Unstructured&CS&.0050&.0050&.0049&.0051&.0073&.0456&.0025&.0059&.0037&.0038&.0033&.0052&.0075&.0109&.0087&.0089&.0106&.0508\\
Unstructured&SE&.0050&.0578&.1345&.1355&.1509&.3878&.0024&.0113&.0057&.0055&.0049&.0033&.0075&.0691&.1402&.1410&.1558&.3911\\
Unstructured&AR1&.0051&.0084&.0267&.0263&.0387&.2379&.0024&.0087&.0043&.0042&.0037&.0032&.0075&.0171&.0310&.0305&.0424&.2412\\
Unstructured&Independent&.0051&.0052&.0065&.0063&.0117&.0988&.0024&.0068&.0038&.0040&.0033&.0045&.0075&.0120&.0103&.0103&.0150&.1033\\
Unstructured&Unstructured&.0051&.0053&.0071&.0075&.0128&.0980&.0024&.0065&.0040&.0040&.0033&.0050&.0075&.0119&.0112&.0115&.0162&.1030\\
\hline
\multicolumn{20}{c}{95th Quantile}\\
\hline
&&\multicolumn{6}{c|}{$ABias^{2}$}&\multicolumn{6}{c|}{$Avar$}&\multicolumn{6}{c}{$AIMSE$}\\
\hline
Structure of $\varepsilon_{X_1}$&Structure of $U_1$&Oracle&SIMEX&FUI&FSMI&Ave&Naive&Oracle&SIMEX&FUI&FSMI&Ave&Naive&Oracle&SIMEX&FUI&FSMI&Ave&Naive\\
\hline
CS&CS&.0031&.0032&.0032&.0036&.0076&.0858&.0097&.0249&.0145&.0136&.0120&.0168&.0128&.0281&.0176&.0172&.0196&.1026\\
CS&SE&.0035&.1303&.2117&.2169&.2312&.4262&.0094&.0360&.0158&.0152&.0131&.0068&.0130&.1663&.2275&.2321&.2443&.4330\\
CS&AR1&.0037&.0132&.0505&.0561&.0715&.3222&.0092&.0325&.0143&.0137&.0118&.0075&.0128&.0458&.0648&.0698&.0834&.3297\\
CS&Independent&.0033&.0035&.0067&.0081&.0169&.1668&.0097&.0278&.0142&.0136&.0117&.0132&.0130&.0313&.0209&.0218&.0286&.1800\\
CS&Unstructured&.0032&.0033&.0043&.0067&.0160&.1610&.0097&.0280&.0163&.0147&.0126&.0157&.0129&.0313&.0206&.0214&.0286&.1767\\
SE&CS&.0039&.0036&.0089&.0087&.0035&.0040&.0024&.0059&.0039&.0038&.0032&.0058&.0063&.0095&.0127&.0125&.0067&.0098\\
SE&SE&.0037&.0043&.0041&.0042&.0084&.0997&.0025&.0232&.0116&.0115&.0096&.0225&.0062&.0275&.0157&.0157&.0181&.1222\\
SE&AR1&.0038&.0034&.0055&.0055&.0035&.0241&.0025&.0102&.0058&.0057&.0048&.0094&.0063&.0136&.0113&.0111&.0083&.0335\\
SE&Independent&.0040&.0041&.0086&.0092&.0037&.0062&.0023&.0066&.0040&.0040&.0033&.0069&.0063&.0107&.0126&.0133&.0071&.0131\\
SE&Unstructured&.0035&.0035&.0116&.0096&.0030&.0047&.0026&.0064&.0049&.0047&.0038&.0094&.0060&.0100&.0165&.0143&.0067&.0142\\
AR1&CS&.0030&.0030&.0061&.0050&.0031&.0085&.0029&.0066&.0042&.0040&.0035&.0060&.0059&.0096&.0103&.0090&.0066&.0144\\
AR1&SE&.0032&.0058&.0250&.0267&.0408&.2507&.0027&.0210&.0102&.0099&.0084&.0099&.0060&.0267&.0351&.0366&.0492&.2606\\
AR1&AR1&.0030&.0029&.0029&.0030&.0068&.0835&.0029&.0121&.0065&.0063&.0054&.0090&.0059&.0150&.0094&.0092&.0122&.0925\\
AR1&Independent&.0029&.0029&.0048&.0043&.0034&.0206&.0029&.0078&.0048&.0047&.0039&.0078&.0058&.0106&.0095&.0090&.0074&.0284\\
AR1&Unstructured&.0028&.0028&.0076&.0050&.0032&.0200&.0030&.0073&.0050&.0046&.0038&.0072&.0058&.0101&.0125&.0096&.0070&.0273\\
Independent&CS&.0035&.0035&.0046&.0131&.0046&.0368&.0062&.0137&.0081&.0093&.0067&.0109&.0097&.0173&.0127&.0225&.0113&.0477\\
Independent&SE&.0035&.0497&.1176&.0971&.1396&.3765&.0063&.0329&.0153&.0179&.0127&.0084&.0098&.0826&.1329&.1150&.1522&.3849\\
Independent&AR1&.0036&.0044&.0147&.0062&.0289&.2266&.0060&.0251&.0119&.0142&.0098&.0091&.0096&.0295&.0266&.0204&.0388&.2357\\
Independent&Independent&.0035&.0034&.0034&.0099&.0074&.0849&.0062&.0182&.0101&.0122&.0084&.0128&.0096&.0216&.0135&.0221&.0158&.0977\\
Independent&Unstructured&.0033&.0032&.0038&.0104&.0069&.0801&.0062&.0163&.0100&.0118&.0077&.0131&.0095&.0194&.0137&.0222&.0147&.0932\\
Unstructured&CS&.0040&.0038&.0040&.0042&.0061&.0455&.0066&.0177&.0095&.0100&.0083&.0132&.0106&.0215&.0136&.0142&.0144&.0587\\
Unstructured&SE&.0042&.0600&.1368&.1386&.1531&.3867&.0068&.0306&.0140&.0142&.0121&.0091&.0109&.0907&.1508&.1528&.1652&.3957\\
Unstructured&AR1&.0041&.0073&.0261&.0258&.0382&.2355&.0068&.0264&.0112&.0113&.0097&.0084&.0108&.0337&.0373&.0371&.0480&.2440\\
Unstructured&Independent&.0042&.0042&.0057&.0054&.0108&.0982&.0066&.0198&.0097&.0098&.0084&.0120&.0108&.0240&.0154&.0152&.0192&.1102\\
Unstructured&Unstructured&.0044&.0049&.0070&.0074&.0128&.0994&.0065&.0199&.0103&.0102&.0085&.0130&.0109&.0248&.0173&.0176&.0213&.1124\\
\hline
\end{longtable}
\end{center}

\begin{table}[h]
\caption{The effect of magnitudes of the functional variable's measurement error on the performance of $\widehat{\beta}_{1\tau}(t)$ obtained under the six methods: oracle, SIMEX, FUI, FSMI, average, and naive considered for Simulation Study 4. The ABias$^2$, Avar, and AIMSE of the estimators for magnitudes of $\sigma_{X_1}(t)=(1.0, 1.5, 2.0, 4.0)$ and $\sigma_{U_1}(t)=(0.5, 1.0, 2.0)$; $\tau=(0.25,0.5,0.75,0.95)$; with $n=500$, AR(1) error structure, $\rho=0.5$, $U_{1}(t)$ was normally distributed, $\sigma_{X_2}=0.5$, $U_{2} \sim N(0, 0.25^{2})$, $\beta_2=0.5$, $\sigma_{Z_c}=0.5$, and $p_{Z_b}=0.6$.}
\label{table: functional_var_ME_full}
\centering
\resizebox{\textwidth}{!}{
\begin{tabular}{rrr|cccccc|cccccc|cccccc}
\hline
\multicolumn{21}{c}{25th Quantile}\\
\hline
&&&\multicolumn{6}{c|}{$ABias^2$}&\multicolumn{6}{c|}{$Avar$}&\multicolumn{6}{c}{$AIMSE$}\\
\hline
$\sigma_{X_1}(t)$&$\sigma_{U_1}(t)$&Ratio&Oracle&SIMEX&FUI&FSMI&Ave&Naive&Oracle&SIMEX&FUI&FSMI&Ave&Naive&Oracle&SIMEX&FUI&FSMI&Ave&Naive\\
\hline
1&0.5&2&.0048&.0047&.0047&.0048&.0053&.0245&.0067&.0146&.0092&.0093&.0085&.0143&.0115&.0192&.0138&.0141&.0139&.0387\\
1&1&1&.0048&.0048&.0048&.0054&.0128&.1287&.0068&.0162&.0109&.0100&.0083&.0100&.0115&.0210&.0156&.0154&.0211&.1387\\
1&2&0.5&.0048&.0140&.0048&.0240&.0711&.3210&.0067&.0182&.0182&.0117&.0073&.0045&.0115&.0322&.0231&.0357&.0784&.3255\\
1.5&0.5&3&.0047&.0046&.0046&.0050&.0047&.0095&.0031&.0068&.0042&.0043&.0041&.0072&.0078&.0114&.0088&.0093&.0088&.0167\\
1.5&1&1.5&.0047&.0047&.0047&.0047&.0064&.0509&.0031&.0074&.0046&.0046&.0041&.0069&.0078&.0120&.0093&.0093&.0105&.0579\\
1.5&2&0.75&.0046&.0053&.0047&.0080&.0251&.2059&.0032&.0100&.0072&.0060&.0046&.0048&.0078&.0152&.0118&.0140&.0297&.2106\\
2&0.5&4&.0046&.0046&.0046&.0051&.0046&.0062&.0019&.0038&.0024&.0025&.0024&.0044&.0065&.0085&.0070&.0076&.0070&.0107\\
2&1&2&.0046&.0046&.0046&.0048&.0052&.0238&.0019&.0043&.0027&.0027&.0025&.0049&.0065&.0089&.0073&.0075&.0077&.0287\\
2&2&1&.0047&.0047&.0047&.0052&.0123&.1273&.0018&.0060&.0040&.0036&.0030&.0046&.0065&.0108&.0086&.0088&.0154&.1318\\
4&0.5&8&.0037&.0037&.0037&.0043&.0037&.0037&.0010&.0016&.0012&.0012&.0012&.0018&.0047&.0053&.0049&.0055&.0049&.0056\\
4&1&4&.0038&.0038&.0037&.0043&.0038&.0053&.0010&.0017&.0012&.0013&.0012&.0021&.0048&.0055&.0050&.0055&.0050&.0074\\
4&2&2&.0039&.0039&.0039&.0040&.0044&.0233&.0009&.0022&.0014&.0015&.0013&.0029&.0049&.0060&.0053&.0055&.0058&.0261\\
\hline
\multicolumn{21}{c}{50th Quantile}\\
\hline
&&&\multicolumn{6}{c|}{$ABias^2$}&\multicolumn{6}{c|}{$Avar$}&\multicolumn{6}{c}{$AIMSE$}\\
\hline
$\sigma_{X_1}(t)$&$\sigma_{U_1}(t)$&Ratio&Oracle&SIMEX&FUI&FSMI&Ave&Naive&Oracle&SIMEX&FUI&FSMI&Ave&Naive&Oracle&SIMEX&FUI&FSMI&Ave&Naive\\
\hline
1&0.5&2&.0047&.0047&.0047&.0048&.0053&.0245&.0060&.0123&.0079&.0079&.0073&.0118&.0107&.0170&.0125&.0127&.0127&.0363\\
1&1&1&.0047&.0047&.0046&.0052&.0125&.1276&.0060&.0134&.0093&.0084&.0071&.0083&.0107&.0181&.0139&.0136&.0196&.1360\\
1&2&0.5&.0048&.0133&.0048&.0235&.0703&.3184&.0060&.0149&.0150&.0096&.0061&.0038&.0107&.0282&.0198&.0331&.0764&.3223\\
1.5&0.5&3&.0047&.0047&.0047&.0050&.0048&.0097&.0028&.0055&.0035&.0036&.0034&.0062&.0074&.0102&.0082&.0086&.0082&.0158\\
1.5&1&1.5&.0047&.0047&.0046&.0047&.0064&.0515&.0028&.0062&.0040&.0039&.0035&.0058&.0074&.0109&.0086&.0086&.0099&.0573\\
1.5&2&0.75&.0047&.0052&.0047&.0079&.0249&.2050&.0028&.0081&.0060&.0050&.0038&.0040&.0075&.0134&.0107&.0129&.0287&.2090\\
2&0.5&4&.0046&.0046&.0046&.0051&.0047&.0064&.0016&.0033&.0021&.0022&.0021&.0037&.0063&.0079&.0067&.0073&.0067&.0100\\
2&1&2&.0046&.0046&.0046&.0047&.0052&.0245&.0016&.0037&.0023&.0023&.0022&.0040&.0063&.0083&.0069&.0070&.0073&.0284\\
2&2&1&.0047&.0047&.0046&.0052&.0123&.1267&.0016&.0050&.0033&.0030&.0025&.0036&.0063&.0096&.0080&.0082&.0148&.1303\\
4&0.5&8&.0040&.0040&.0040&.0045&.0040&.0041&.0009&.0013&.0010&.0010&.0010&.0014&.0048&.0052&.0049&.0055&.0049&.0055\\
4&1&4&.0041&.0041&.0041&.0046&.0041&.0058&.0008&.0013&.0009&.0010&.0009&.0016&.0049&.0054&.0050&.0056&.0051&.0074\\
4&2&2&.0041&.0041&.0041&.0043&.0047&.0236&.0008&.0017&.0012&.0012&.0011&.0022&.0049&.0059&.0053&.0055&.0058&.0257\\
\hline
\multicolumn{21}{c}{75th Quantile}\\
\hline
&&&\multicolumn{6}{c|}{$ABias^2$}&\multicolumn{6}{c|}{$Avar$}&\multicolumn{6}{c}{$AIMSE$}\\
\hline
$\sigma_{X_1}(t)$&$\sigma_{U_1}(t)$&Ratio&Oracle&SIMEX&FUI&FSMI&Ave&Naive&Oracle&SIMEX&FUI&FSMI&Ave&Naive&Oracle&SIMEX&FUI&FSMI&Ave&Naive\\
\hline
1&0.5&2&.0047&.0048&.0048&.0048&.0056&.0257&.0067&.0142&.0089&.0091&.0083&.0132&.0115&.0190&.0137&.0139&.0138&.0389\\
1&1&1&.0046&.0047&.0046&.0052&.0127&.1272&.0068&.0169&.0113&.0102&.0086&.0100&.0114&.0216&.0159&.0155&.0213&.1372\\
1&2&0.5&.0046&.0131&.0046&.0230&.0695&.3187&.0068&.0190&.0191&.0122&.0077&.0045&.0114&.0321&.0238&.0352&.0773&.3231\\
1.5&0.5&3&.0046&.0047&.0046&.0049&.0048&.0099&.0032&.0066&.0041&.0043&.0040&.0070&.0079&.0112&.0088&.0092&.0088&.0170\\
1.5&1&1.5&.0046&.0046&.0046&.0046&.0064&.0514&.0033&.0076&.0048&.0047&.0042&.0069&.0078&.0122&.0094&.0093&.0106&.0583\\
1.5&2&0.75&.0046&.0051&.0046&.0077&.0247&.2036&.0032&.0101&.0073&.0061&.0046&.0048&.0078&.0152&.0119&.0138&.0294&.2084\\
2&0.5&4&.0044&.0044&.0044&.0048&.0045&.0063&.0019&.0038&.0024&.0025&.0024&.0043&.0064&.0082&.0068&.0073&.0069&.0106\\
2&1&2&.0044&.0044&.0044&.0045&.0050&.0241&.0019&.0045&.0028&.0028&.0026&.0048&.0064&.0089&.0072&.0073&.0076&.0289\\
2&2&1&.0045&.0046&.0045&.0051&.0121&.1263&.0019&.0060&.0039&.0036&.0030&.0046&.0064&.0106&.0085&.0087&.0151&.1309\\
4&0.5&8&.0037&.0037&.0037&.0042&.0037&.0038&.0010&.0015&.0011&.0012&.0011&.0018&.0047&.0052&.0048&.0054&.0048&.0055\\
4&1&4&.0036&.0036&.0036&.0040&.0036&.0053&.0011&.0018&.0013&.0014&.0013&.0021&.0047&.0053&.0049&.0054&.0049&.0074\\
4&2&2&.0036&.0036&.0036&.0037&.0042&.0228&.0011&.0023&.0016&.0016&.0015&.0029&.0047&.0059&.0052&.0054&.0057&.0257\\
\hline
\multicolumn{21}{c}{95th Quantile}\\
\hline
&&&\multicolumn{6}{c|}{$ABias^2$}&\multicolumn{6}{c|}{$Avar$}&\multicolumn{6}{c}{$AIMSE$}\\
\hline
$\sigma_{X_1}(t)$&$\sigma_{U_1}(t)$&Ratio&Oracle&SIMEX&FUI&FSMI&Ave&Naive&Oracle&SIMEX&FUI&FSMI&Ave&Naive&Oracle&SIMEX&FUI&FSMI&Ave&Naive\\
\hline
1&0.5&2&.0035&.0038&.0036&.0036&.0045&.0237&.0179&.0420&.0233&.0239&.0217&.0357&.0214&.0458&.0269&.0274&.0262&.0594\\
1&1&1&.0035&.0037&.0035&.0044&.0122&.1268&.0178&.0474&.0278&.0259&.0213&.0248&.0213&.0512&.0313&.0302&.0335&.1516\\
1&2&0.5&.0038&.0132&.0035&.0232&.0707&.3168&.0176&.0522&.0470&.0308&.0190&.0116&.0213&.0654&.0505&.0540&.0897&.3283\\
1.5&0.5&3&.0035&.0035&.0034&.0036&.0036&.0089&.0085&.0191&.0106&.0111&.0103&.0191&.0119&.0226&.0140&.0148&.0139&.0280\\
1.5&1&1.5&.0035&.0036&.0035&.0035&.0055&.0503&.0087&.0223&.0123&.0122&.0108&.0184&.0122&.0259&.0158&.0157&.0163&.0687\\
1.5&2&0.75&.0035&.0043&.0034&.0070&.0248&.2026&.0087&.0310&.0196&.0163&.0124&.0119&.0122&.0353&.0230&.0233&.0372&.2145\\
2&0.5&4&.0033&.0034&.0033&.0036&.0034&.0048&.0053&.0114&.0064&.0067&.0063&.0115&.0086&.0148&.0097&.0103&.0097&.0163\\
2&1&2&.0033&.0033&.0032&.0033&.0039&.0229&.0053&.0132&.0072&.0074&.0067&.0121&.0086&.0164&.0105&.0107&.0107&.0350\\
2&2&1&.0033&.0033&.0032&.0038&.0111&.1232&.0052&.0189&.0105&.0099&.0081&.0111&.0085&.0222&.0137&.0137&.0192&.1343\\
4&0.5&8&.0024&.0025&.0024&.0029&.0024&.0025&.0022&.0039&.0026&.0026&.0025&.0042&.0047&.0064&.0050&.0056&.0050&.0067\\
4&1&4&.0023&.0023&.0023&.0027&.0023&.0039&.0023&.0045&.0028&.0029&.0028&.0051&.0046&.0068&.0051&.0056&.0051&.0090\\
4&2&2&.0024&.0023&.0023&.0025&.0029&.0221&.0022&.0064&.0036&.0037&.0033&.0070&.0046&.0087&.0059&.0062&.0063&.0291\\
\hline
\end{tabular}}
\end{table}

\begin{table}[h]
\caption{The effect of magnitudes of the scalar variable's measurement error on the performance of $\widehat{\beta}_{1\tau}(t)$ obtained under the six methods: oracle, SIMEX, FUI, FSMI, average, and naive considered for Simulation Study 4. The ABias$^2$, Avar, and AIMSE of the estimators for magnitudes of $\sigma_{X_2}=(1.0, 1.5, 2.0, 4.0)$ and $\sigma_{U_2}=(0.5, 1.0, 2.0)$; $\tau=(0.25,0.5,0.75,0.95)$; with $n=500$, AR(1) error structure, $\rho=0.5$, $\sigma_{X_1}(t)=3$, $U_{1}(t) \sim MVN(0, 2.5^2)$, $U_{2}$ was normally distributed, $\beta_2=0.5$, $\sigma_{Z_c}=0.5$, and $p_{Z_b}=0.6$.}
\label{table: scalar_var_ME_full}
\centering
\resizebox{\textwidth}{!}{
\begin{tabular}{rrr|cccccc|cccccc|cccccc}
\hline
\multicolumn{21}{c}{25th Quantile}\\
\hline
&&&\multicolumn{6}{c|}{$ABias^2$}&\multicolumn{6}{c|}{$Avar$}&\multicolumn{6}{c}{$AIMSE$}\\
\hline
$\sigma_{X_2}$&$\sigma_{U_2}$&Ratio&Oracle&SIMEX&FUI&FSMI&Ave&Naive&Oracle&SIMEX&FUI&FSMI&Ave&Naive&Oracle&SIMEX&FUI&FSMI&Ave&Naive\\
\hline
1&0.5&2&.0046&.0045&.0045&.0046&.0085&.0872&.0009&.0044&.0028&.0027&.0023&.0053&.0055&.0089&.0073&.0073&.0108&.0925\\
1&1&1&.0046&.0045&.0045&.0046&.0086&.0871&.0009&.0080&.0049&.0046&.0040&.0084&.0055&.0125&.0093&.0092&.0127&.0955\\
1&2&0.5&.0046&.0046&.0045&.0046&.0086&.0861&.0009&.0167&.0099&.0096&.0082&.0115&.0055&.0213&.0144&.0141&.0168&.0976\\
1.5&0.5&3&.0046&.0045&.0046&.0046&.0085&.0872&.0009&.0045&.0028&.0027&.0023&.0056&.0055&.0090&.0073&.0073&.0108&.0928\\
1.5&1&1.5&.0046&.0045&.0045&.0046&.0086&.0870&.0009&.0081&.0050&.0047&.0041&.0110&.0055&.0127&.0095&.0093&.0127&.0981\\
1.5&2&0.75&.0046&.0046&.0045&.0047&.0089&.0871&.0009&.0199&.0119&.0114&.0098&.0183&.0055&.0246&.0164&.0160&.0187&.1055\\
2&0.5&4&.0047&.0047&.0047&.0048&.0087&.0870&.0009&.0044&.0027&.0026&.0022&.0056&.0056&.0090&.0074&.0074&.0109&.0926\\
2&1&2&.0047&.0047&.0047&.0047&.0087&.0875&.0009&.0081&.0049&.0047&.0040&.0122&.0056&.0128&.0096&.0094&.0128&.0997\\
2&2&1&.0047&.0048&.0047&.0048&.0090&.0871&.0009&.0214&.0127&.0121&.0106&.0243&.0056&.0261&.0174&.0169&.0195&.1113\\
4&0.5&8&.0047&.0047&.0047&.0048&.0087&.0866&.0009&.0044&.0027&.0026&.0022&.0057&.0056&.0090&.0074&.0074&.0109&.0923\\
4&1&4&.0047&.0047&.0047&.0048&.0087&.0876&.0009&.0083&.0050&.0049&.0042&.0135&.0056&.0129&.0097&.0096&.0129&.1010\\
4&2&2&.0047&.0047&.0047&.0048&.0090&.0889&.0009&.0225&.0136&.0131&.0112&.0390&.0056&.0272&.0182&.0179&.0202&.1280\\
\hline
\multicolumn{21}{c}{50th Quantile}\\
\hline
&&&\multicolumn{6}{c|}{$ABias^2$}&\multicolumn{6}{c|}{$Avar$}&\multicolumn{6}{c}{$AIMSE$}\\
\hline
$\sigma_{X_2}$&$\sigma_{U_2}$&Ratio&Oracle&SIMEX&FUI&FSMI&Ave&Naive&Oracle&SIMEX&FUI&FSMI&Ave&Naive&Oracle&SIMEX&FUI&FSMI&Ave&Naive\\
\hline
1&0.5&2&.0046&.0046&.0046&.0047&.0086&.0867&.0008&.0037&.0024&.0023&.0020&.0044&.0055&.0083&.0070&.0070&.0105&.0911\\
1&1&1&.0046&.0047&.0046&.0047&.0087&.0867&.0008&.0065&.0041&.0039&.0034&.0069&.0055&.0112&.0088&.0086&.0121&.0936\\
1&2&0.5&.0046&.0047&.0046&.0047&.0088&.0868&.0008&.0131&.0082&.0077&.0067&.0094&.0055&.0178&.0128&.0125&.0156&.0962\\
1.5&0.5&3&.0047&.0047&.0047&.0047&.0086&.0864&.0008&.0037&.0023&.0022&.0019&.0046&.0055&.0083&.0070&.0070&.0105&.0910\\
1.5&1&1.5&.0047&.0047&.0047&.0048&.0087&.0862&.0008&.0067&.0043&.0041&.0035&.0086&.0055&.0114&.0089&.0089&.0122&.0948\\
1.5&2&0.75&.0047&.0047&.0047&.0048&.0088&.0869&.0008&.0160&.0097&.0093&.0081&.0149&.0055&.0207&.0144&.0140&.0168&.1018\\
2&0.5&4&.0047&.0047&.0047&.0047&.0086&.0867&.0008&.0037&.0023&.0022&.0019&.0047&.0055&.0083&.0070&.0070&.0105&.0914\\
2&1&2&.0047&.0047&.0047&.0048&.0087&.0865&.0008&.0067&.0043&.0041&.0036&.0095&.0055&.0114&.0090&.0089&.0123&.0960\\
2&2&1&.0047&.0047&.0046&.0048&.0088&.0864&.0008&.0168&.0103&.0099&.0085&.0205&.0055&.0215&.0149&.0146&.0173&.1069\\
4&0.5&8&.0047&.0047&.0047&.0048&.0086&.0866&.0008&.0036&.0023&.0022&.0019&.0048&.0055&.0083&.0070&.0070&.0105&.0914\\
4&1&4&.0047&.0047&.0047&.0048&.0088&.0862&.0008&.0066&.0042&.0040&.0035&.0110&.0055&.0113&.0089&.0089&.0123&.0972\\
4&2&2&.0047&.0047&.0047&.0048&.0089&.0861&.0008&.0181&.0115&.0111&.0095&.0315&.0055&.0228&.0162&.0159&.0184&.1176\\
\hline
\multicolumn{21}{c}{75th Quantile}\\
\hline
&&&\multicolumn{6}{c|}{$ABias^2$}&\multicolumn{6}{c|}{$Avar$}&\multicolumn{6}{c}{$AIMSE$}\\
\hline
$\sigma_{X_2}$&$\sigma_{U_2}$&Ratio&Oracle&SIMEX&FUI&FSMI&Ave&Naive&Oracle&SIMEX&FUI&FSMI&Ave&Naive&Oracle&SIMEX&FUI&FSMI&Ave&Naive\\
\hline
1&0.5&2&.0044&.0044&.0043&.0044&.0084&.0859&.0011&.0045&.0029&.0027&.0024&.0054&.0054&.0089&.0072&.0072&.0108&.0914\\
1&1&1&.0044&.0043&.0043&.0044&.0086&.0862&.0011&.0077&.0049&.0047&.0040&.0085&.0054&.0121&.0092&.0091&.0126&.0947\\
1&2&0.5&.0044&.0043&.0042&.0044&.0087&.0867&.0011&.0171&.0103&.0099&.0085&.0122&.0054&.0213&.0145&.0143&.0173&.0989\\
1.5&0.5&3&.0046&.0045&.0045&.0046&.0085&.0864&.0010&.0045&.0028&.0026&.0023&.0057&.0055&.0090&.0073&.0072&.0108&.0921\\
1.5&1&1.5&.0046&.0046&.0045&.0046&.0087&.0874&.0010&.0077&.0048&.0046&.0040&.0107&.0055&.0123&.0093&.0092&.0127&.0981\\
1.5&2&0.75&.0046&.0045&.0044&.0046&.0089&.0882&.0010&.0198&.0119&.0114&.0099&.0186&.0055&.0244&.0163&.0160&.0188&.1068\\
2&0.5&4&.0046&.0046&.0046&.0047&.0086&.0865&.0009&.0044&.0027&.0026&.0023&.0059&.0056&.0090&.0073&.0073&.0108&.0924\\
2&1&2&.0046&.0046&.0046&.0047&.0088&.0883&.0009&.0077&.0047&.0045&.0039&.0121&.0056&.0123&.0093&.0092&.0127&.1004\\
2&2&1&.0046&.0046&.0046&.0048&.0091&.0904&.0009&.0209&.0128&.0121&.0106&.0247&.0056&.0255&.0174&.0169&.0197&.1151\\
4&0.5&8&.0047&.0047&.0047&.0047&.0086&.0862&.0009&.0044&.0027&.0026&.0022&.0060&.0056&.0090&.0074&.0073&.0109&.0922\\
4&1&4&.0047&.0047&.0047&.0048&.0089&.0868&.0009&.0077&.0047&.0045&.0039&.0137&.0056&.0124&.0094&.0093&.0128&.1005\\
4&2&2&.0047&.0048&.0047&.0049&.0091&.0879&.0009&.0217&.0134&.0127&.0111&.0384&.0056&.0265&.0181&.0176&.0202&.1263\\
\hline
\multicolumn{21}{c}{95th Quantile}\\
\hline
&&&\multicolumn{6}{c|}{$ABias^2$}&\multicolumn{6}{c|}{$Avar$}&\multicolumn{6}{c}{$AIMSE$}\\
\hline
$\sigma_{X_2}$&$\sigma_{U_2}$&Ratio&Oracle&SIMEX&FUI&FSMI&Ave&Naive&Oracle&SIMEX&FUI&FSMI&Ave&Naive&Oracle&SIMEX&FUI&FSMI&Ave&Naive\\
\hline
1&0.5&2&.0034&.0033&.0033&.0034&.0073&.0857&.0027&.0146&.0076&.0073&.0063&.0132&.0061&.0179&.0109&.0107&.0136&.0989\\
1&1&1&.0034&.0034&.0033&.0034&.0073&.0844&.0027&.0238&.0123&.0119&.0102&.0222&.0061&.0272&.0157&.0153&.0175&.1065\\
1&2&0.5&.0034&.0036&.0033&.0035&.0078&.0859&.0027&.0486&.0249&.0245&.0206&.0309&.0061&.0522&.0283&.0280&.0284&.1168\\
1.5&0.5&3&.0037&.0037&.0037&.0038&.0077&.0856&.0026&.0149&.0076&.0075&.0063&.0137&.0063&.0186&.0113&.0112&.0140&.0993\\
1.5&1&1.5&.0037&.0037&.0037&.0037&.0076&.0832&.0026&.0250&.0128&.0125&.0106&.0276&.0063&.0287&.0165&.0162&.0182&.1108\\
1.5&2&0.75&.0037&.0037&.0035&.0036&.0078&.0830&.0026&.0592&.0303&.0292&.0250&.0492&.0063&.0629&.0337&.0328&.0328&.1322\\
2&0.5&4&.0037&.0036&.0036&.0037&.0077&.0861&.0026&.0147&.0077&.0075&.0063&.0143&.0063&.0183&.0113&.0112&.0140&.1003\\
2&1&2&.0037&.0035&.0035&.0036&.0075&.0853&.0026&.0250&.0129&.0126&.0107&.0319&.0063&.0286&.0165&.0162&.0182&.1172\\
2&2&1&.0037&.0034&.0034&.0037&.0078&.0862&.0026&.0636&.0318&.0306&.0262&.0650&.0063&.0670&.0352&.0342&.0340&.1512\\
4&0.5&8&.0036&.0035&.0035&.0036&.0074&.0855&.0026&.0151&.0080&.0076&.0066&.0143&.0062&.0186&.0115&.0112&.0140&.0999\\
4&1&4&.0036&.0034&.0034&.0035&.0074&.0861&.0026&.0260&.0131&.0129&.0109&.0352&.0062&.0294&.0165&.0164&.0183&.1213\\
4&2&2&.0036&.0032&.0032&.0033&.0075&.0873&.0026&.0699&.0354&.0345&.0293&.1006&.0062&.0731&.0386&.0378&.0368&.1879\\
\hline
\end{tabular}}
\end{table}

\begin{table}[h]
\caption{The effect of $\beta_2$ on the performance of $\widehat{\beta}_{1\tau}(t)$ obtained under the six methods: oracle, SIMEX, FUI, FSMI, average, and naive considered for Simulation Study 5. The ABias$^2$, Avar, and AIMSE of the estimators for $\beta_2=(0.5, 1.0, 1.5, 2.0, 4.0)$; $\tau=(0.25,0.5,0.75,0.95)$; with $n=500$, AR(1) error structure, $\rho=0.5$, $\sigma_{X_1}(t)=3$, $U_{1}(t) \sim MVN(0, 2.5^2)$, $\sigma_{X_2}=0.5$, $U_{2} \sim N(0, 0.25^{2})$, $\sigma_{Z_c}=0.5$, and $p_{Z_b}=0.6$.}
\label{tabe: beta2_full}
\centering
\resizebox{\textwidth}{!}{
\begin{tabular}{r|cccccc|cccccc|cccccc}
\hline
\multicolumn{19}{c}{25th Quantile}\\
\hline
&\multicolumn{6}{c|}{$ABias^2$}&\multicolumn{6}{c|}{$Avar$}&\multicolumn{6}{c}{$AIMSE$}\\
\hline
$\beta_{2}$&Oracle&SIMEX&FUI&FSMI&Ave&Naive&Oracle&SIMEX&FUI&FSMI&Ave&Naive&Oracle&SIMEX&FUI&FSMI&Ave&Naive\\
\hline
0.5&.0043&.0043&.0043&.0044&.0083&.0861&.0011&.0036&.0023&.0023&.0019&.0037&.0054&.0080&.0066&.0066&.0102&.0898\\
1&.0046&.0045&.0045&.0046&.0085&.0872&.0009&.0044&.0028&.0027&.0023&.0053&.0055&.0089&.0073&.0073&.0108&.0925\\
1.5&.0046&.0045&.0045&.0046&.0086&.0876&.0009&.0062&.0038&.0036&.0032&.0083&.0055&.0107&.0084&.0083&.0118&.0959\\
2&.0047&.0047&.0047&.0047&.0087&.0875&.0009&.0081&.0049&.0047&.0040&.0122&.0056&.0128&.0096&.0094&.0128&.0997\\
4&.0047&.0047&.0047&.0048&.0090&.0889&.0009&.0225&.0136&.0131&.0112&.0390&.0056&.0272&.0182&.0179&.0202&.1280\\
\hline
\multicolumn{19}{c}{50th Quantile}\\
\hline
&\multicolumn{6}{c|}{$ABias^2$}&\multicolumn{6}{c|}{$Avar$}&\multicolumn{6}{c}{$AIMSE$}\\
\hline
$\beta_{2}$&Oracle&SIMEX&FUI&FSMI&Ave&Naive&Oracle&SIMEX&FUI&FSMI&Ave&Naive&Oracle&SIMEX&FUI&FSMI&Ave&Naive\\
\hline
0.5&.0045&.0044&.0044&.0045&.0083&.0868&.0009&.0031&.0020&.0019&.0017&.0029&.0054&.0075&.0065&.0064&.0100&.0897\\
1&.0046&.0046&.0046&.0047&.0086&.0867&.0008&.0037&.0024&.0023&.0020&.0044&.0055&.0083&.0070&.0070&.0105&.0911\\
1.5&.0047&.0047&.0047&.0047&.0087&.0860&.0008&.0050&.0031&.0030&.0026&.0065&.0055&.0096&.0078&.0077&.0112&.0925\\
2&.0047&.0047&.0047&.0048&.0087&.0865&.0008&.0067&.0043&.0041&.0036&.0095&.0055&.0114&.0090&.0089&.0123&.0960\\
4&.0047&.0047&.0047&.0048&.0089&.0861&.0008&.0181&.0115&.0111&.0095&.0315&.0055&.0228&.0162&.0159&.0184&.1176\\
\hline
\multicolumn{19}{c}{75th Quantile}\\
\hline
&\multicolumn{6}{c|}{$ABias^2$}&\multicolumn{6}{c|}{$Avar$}&\multicolumn{6}{c}{$AIMSE$}\\
\hline
$\beta_{2}$&Oracle&SIMEX&FUI&FSMI&Ave&Naive&Oracle&SIMEX&FUI&FSMI&Ave&Naive&Oracle&SIMEX&FUI&FSMI&Ave&Naive\\
\hline
0.5&.0043&.0043&.0042&.0043&.0082&.0854&.0011&.0037&.0024&.0023&.0020&.0038&.0054&.0079&.0066&.0066&.0101&.0892\\
1&.0044&.0044&.0043&.0044&.0084&.0859&.0011&.0045&.0029&.0027&.0024&.0054&.0054&.0089&.0072&.0072&.0108&.0914\\
1.5&.0046&.0045&.0045&.0046&.0086&.0875&.0010&.0059&.0036&.0035&.0030&.0082&.0055&.0104&.0081&.0081&.0116&.0957\\
2&.0046&.0046&.0046&.0047&.0088&.0883&.0009&.0077&.0047&.0045&.0039&.0121&.0056&.0123&.0093&.0092&.0127&.1004\\
4&.0047&.0048&.0047&.0049&.0091&.0879&.0009&.0217&.0134&.0127&.0111&.0384&.0056&.0265&.0181&.0176&.0202&.1263\\
\hline
\multicolumn{19}{c}{95th Quantile}\\
\hline
&\multicolumn{6}{c|}{$ABias^2$}&\multicolumn{6}{c|}{$Avar$}&\multicolumn{6}{c}{$AIMSE$}\\
\hline
$\beta_{2}$&Oracle&SIMEX&FUI&FSMI&Ave&Naive&Oracle&SIMEX&FUI&FSMI&Ave&Naive&Oracle&SIMEX&FUI&FSMI&Ave&Naive\\
\hline
0.5&.0030&.0029&.0029&.0030&.0068&.0835&.0029&.0121&.0065&.0063&.0054&.0090&.0059&.0150&.0094&.0092&.0122&.0925\\
1&.0034&.0033&.0033&.0034&.0073&.0857&.0027&.0146&.0076&.0073&.0063&.0132&.0061&.0179&.0109&.0107&.0136&.0989\\
1.5&.0037&.0037&.0037&.0038&.0077&.0838&.0026&.0193&.0099&.0095&.0082&.0205&.0063&.0230&.0136&.0133&.0159&.1044\\
2&.0037&.0035&.0035&.0036&.0075&.0853&.0026&.0250&.0129&.0126&.0107&.0319&.0063&.0286&.0165&.0162&.0182&.1172\\
4&.0036&.0032&.0032&.0033&.0075&.0873&.0026&.0699&.0354&.0345&.0293&.1006&.0062&.0731&.0386&.0378&.0368&.1879\\
\hline
\end{tabular}}
\end{table}